# Spontaneous chiral symmetry breaking in polar fluid – heliconical ferroelectric nematic phase

Jakub Karcz[1], Jakub Herman[1], Natan Rychłowicz[1], Przemysław Kula[1]*, Ewa Górecka[2], Jadwiga Szydlowska[2], Pawel W. Majewski[2], Damian Pociecha[2]*

[1]Faculty of Advanced Technology and Chemistry, Military University of Technology; Warsaw, Poland.

[2]Faculty of Chemistry, University of Warsaw; Warsaw, Poland.

*Corresponding authors: przemyslaw.kula@wat.edu.pl, pociu@chem.uw.edu.pl

Spontaneous mirror symmetry breaking by formation of chiral structures from achiral building blocks and emergent polar order are phenomena rarely observed in fluids. Separately, they have been both found in certain nematic liquid crystalline phases, however, they have never been observed simultaneously. Here, we report a heliconical arrangement of achiral molecules in the ferroelectric nematic phase. The phase is thus spontaneously both polar and chiral. Notably, the pitch of the heliconical structure is comparable to the wavelength of visible light giving selective reflection controllable by temperature or application of a weak electric field. Despite bearing resemblance to the heliconical twist-bend nematic phase, this chiral ferroelectric nematic phase arises from electrical interactions that induce a non-collinear orientation of electric dipoles.

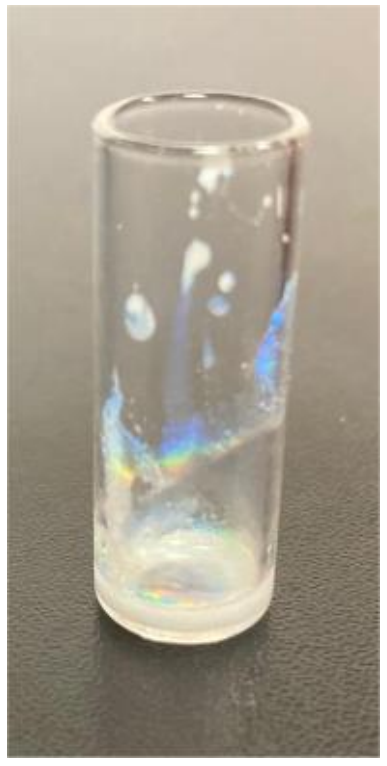

In conventional nematic, N, liquid crystals (LC) anisotropic molecules align along a common direction called the *director*, but lack long-range positional order. When built from chiral molecules, the nematic phase exhibits a helical structure, with the helix axis perpendicular to the director. The nematic phase has been the longest-studied liquid crystalline phase, often regarded as the most useful for practical applications, but lacking ‘scientific novelty’. Therefore, the recent discoveries of two variants of nematic phase: twist-bend, $N_{TB}$, (*1 – 3*) and ferroelectric, $N_F$, (*4 – 6*) were very surprising. The $N_{TB}$ phase, although it is formed by achiral molecules, exhibits a heliconical structure, in which the molecules are tilted at an arbitrary angle to the helical axis and its helical pitch is remarkably short, spanning just a few molecular distances. In the ferroelectric nematic phase, the constituent molecular dipoles align along the *director*, resulting in a high spontaneous electric polarization, comparable to that of solid crystalline ferroelectrics. Its discovery challenged the common belief that dipole interactions in liquids are too weak, and thermal fluctuations are too strong, to sustain the long-range ordering of electric dipole moments. The polar order has been previously observed in liquid crystalline systems with some degree of positional order, it aroused as a secondary effect driven by steric interactions: molecular chirality (*7*) or constrained rotations of bent-core molecules (*8*). In the $N_F$ phase, the dipole order is exceptionally strong, highlighting the unexpected robustness of dipole interactions. Moreover, electric dipoles exhibit a strong preference for co-linearity, with the $N_F$ phase being characterized by a ground state with uniform polar order. To induce the twist of dipoles in the polar nematic phase, akin to the non-polar chiral nematic phase, chiral mesogens were used (*9*), or the liquid crystalline material was doped with chiral additives (*10 - 12*). In such cases, the helical arrangement of molecules also induces a helical arrangement of local polarization vectors, perpendicular to the helical axis. However, for achiral rod-like molecules, no inherent mechanism facilitating the spontaneous twisting of dipoles in the nematic phase has been reported. The question of whether dipole interactions may lead to spontaneous mirror symmetry breaking and the formation of twisted structures remains open. Such a mechanism, Dzyaloshinskii-Moriya (DM) interaction, is well known for magnetic systems (*13*) and results in a rich plethora of structures, including helices, skyrmions, and merons, (*14*) wherein magnetic dipoles spontaneously align into twisted arrangements. Chiral symmetry breaking by electric dipole interactions would be a captivating phenomenon wherein an originally achiral polar system spontaneously adopts either a left-handed or right-handed polar helical configuration. The twisted dipoles could form a helical structure analogous to the nematic phase made of chiral molecules, in which the bulk polarization would be fully compensated, or they could form an axially polar heliconical structure, with the polarization only partially compensated, a structure corresponding to the twist-bend nematic phase.

## Materials

4'-(Difluoro(3,4,5-trifluorophenoxy)methyl)-2,3',5'-trifluoro-[1,1'-biphenyl]-4-yl 2,6-difluoro-4-(5-propyl-1,3-dioxan-2-yl)benzoate (synthesis described in SM, Fig. S1), hereafter referred to as **MUT_JK103**, was synthesized in the Liquid Crystal group of the Military University of Technology; its molecular structure is based on DIO – one of the model ferronematogens (*4*). The mesogenic core was elongated; in the optimized geometry (Fig. S2) the molecular length is 31.9 Å. The dipole moment, being nearly collinear with the long molecular axis (the calculated angle between dipole direction and long molecular axis is 9.5°), was increased to 12.6 D by substituting additional fluorine atoms in the core. The compound **MUT_JK103** forms 5 mesophases between the isotropic liquid and the solid crystal: N – conventional nematic, $N_x$ (also known as $SmZ_A$) – nematic with a periodic antiferroelectric domain structure (*15, 16*), $N_F$ – ferroelectric nematic, $N_{TBF}$ – ferroelectric twist-bent type nematic and $SmC_F$ – tilted ferroelectric smectic (transition temperatures are given in SM). It should be emphasized that all LC phases are enantiotropic, with small but distinct thermal effects accompanying the phase transitions (Fig. 1).

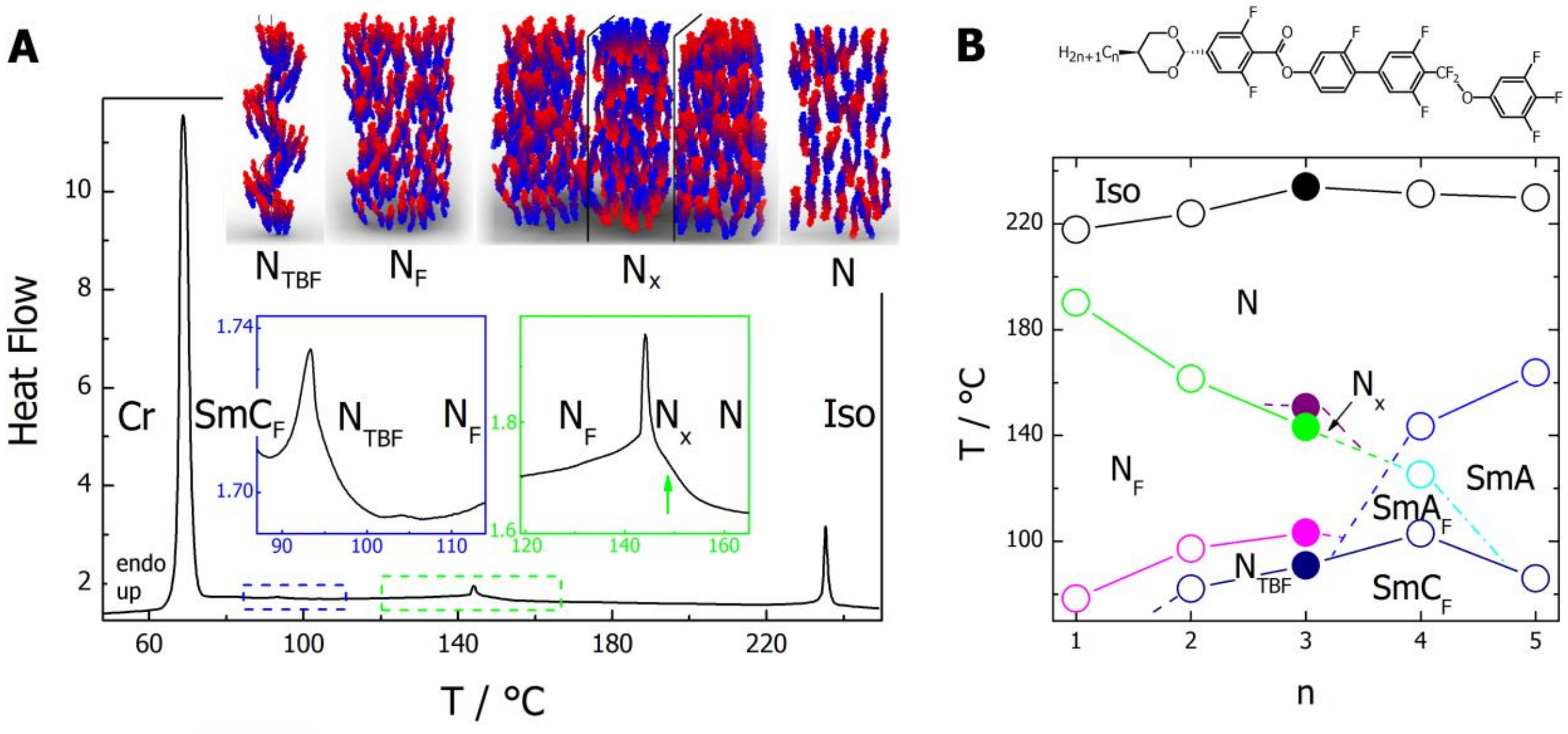

**Fig. 1. Studied material and phase sequence. A** DSC thermogram recorded on second heating scan for **MUT_JK103**, with enlarged temperature ranges of transitions between LC phases, note that all the LC phases appear above the melting of the solid crystal phase, Cr. In the inset, schematic drawings showing molecular organization in various nematic phases, the red and blue shading of each molecule shows negative and positive ends of their dipole moments. All phases except the N phase lack a head-tail orientation equivalence, thus polar order occurs; in the $N_x$ phase regular antiferroelectric structure of uniformly polar domains is observed, the $N_F$ phase shows global ferroelectric order, and in $N_{TBF}$ phase molecules form heliconical structure, with effective polarization along the helical axis. **B** Phase diagram for the homolog series **MUT_JK10n**, the general molecular structure is presented above. Filled symbols refer to the homolog n = 3, which is studied in detail. $SmA_F$ and $SmC_F$ refer to lamellar phases, orthogonal and tilted, respectively, with ferroelectric order.

## X-ray diffraction

X-ray diffraction studies (Fig. S3) confirmed that four of the mesophases lack long-range positional order of molecules, so they are identified as nematic phases. The width of the low angle diffraction signal gradually decreases with decreasing temperature which corresponds to an increase of positional correlation length. The correlation length reaches ~100 Å, (~ 3 molecular lengths) in the lowest temperature nematic phase. On further cooling the signal narrows to be limited by instrumental broadening only, evidencing one-dimensional long range positional order, characteristic for smectics. The diffuse high angle diffraction signal shows a liquid-like order of molecules within the smectic layers. The position of the low angle diffraction signal in the N, $N_x$ and $N_F$ phases is almost independent of temperature. In the $N_{TBF}$ phase it gradually moves to higher angles indicating a decrease in periodicity. The trend continues in the smectic phase, the measured periodicity decreases due to tilting of molecules – therefore the phase was identified as SmC - lamellar phase with molecules tilted with respect to the layer normal and with short-range positional order within the layers.

## Polar properties

The phases $N_F$, $N_{TBF}$ and $SmC_F$ are polar. When a weak electric field (200 mV $\mu m^{-1}$) is applied along the cell thickness, the planar optical texture transforms into a homeotropic, non-birefringent one as molecules realign with their dipole moments along the electric field. Moreover, a polarization hysteresis loop is observed, as well as switching current peak, when a triangular-wave voltage is applied (Fig. S4). The corresponding spontaneous electric polarization is of the order of 4 $\mu C\ cm^{-2}$, a typical value for other $N_F$ materials (*17*). Dielectric spectroscopy revealed a strong collective relaxation mode in the kHz range in the polar nematic phases (Fig. S5), with a static permittivity of the order of $10^4$, the mode is also visible in the $SmC_F$ phase, however its strength gradually decreases.

## Optical textures

Optical studies using polarized light optical microscopy (Fig. 2 and Fig. S6) showed that the highest temperature LC phase exhibits textures characteristic of a conventional nematic, i.e., a uniform texture in a cell with planar anchoring, with the *director* aligned along the rubbing direction. Strong flickering due to director fluctuations was observed. The transition to the lower-temperature nematic phase was evidenced by the cessation of flickering and the appearance of chevron defects typical of the $N_x$ phase (*15*). In the subsequent $N_F$ phase, in a cell with parallel rubbing on both surfaces, the uniform texture reappears with the optical axis along the rubbing direction, with some parabolic defects (*18*) anchored to the plastic beads used as cell spacers. In a cell with antiparallel rubbing optically active domains in which the director twists between the surfaces are observed (Fig. S7). Such twisted domains were reported as characteristic of the $N_F$ phase (*19*). Finally, in the $N_{TBF}$ phase, on cooling a few K below the phase transition, stripes begin to form along the rubbing direction. The periodicity of the stripes is independent of temperature but varies with cell thickness; it is 3 μm for a 1.5-μm-thick cell and for thicker cells the periodicity is larger. In thicker samples the stripes usually nucleate from defects, following their formation it was deduced that neighboring stripes have opposite polarization directions (Fig. S8). In the smectic phase, the texture is grainy, uncharacteristic, and strongly light-scattering.

## Fiber formation

The investigated material in the $N_F$ and $N_{TBF}$ phases readily forms cylindrical fibers in the form of bridges between the bulk material. These freely suspended strands between two bulky ends are of nearly constant thickness (Fig. 2CD). In the $N_F$ phase, the fibers usually rupture when their aspect ratio (AR, length/diameter) approaches 70, which is significantly larger than predicted by Rayleigh-Plateau theory for the isotropic phase. They are stabilized by a polar order that suppresses the fluctuation of the director (*20, 21*). In the $N_{TBF}$ phase, the fibers are thinner, and their aspect ratio easily reaches 100 before they break, which suggests an additional mechanism responsible for their exceptional stability. Such a mechanism is usually related to the internal structure of the phase, e.g. for smectic and columnar phases it was attributed to the local stiffness of the layers/columns against compression (*22*).

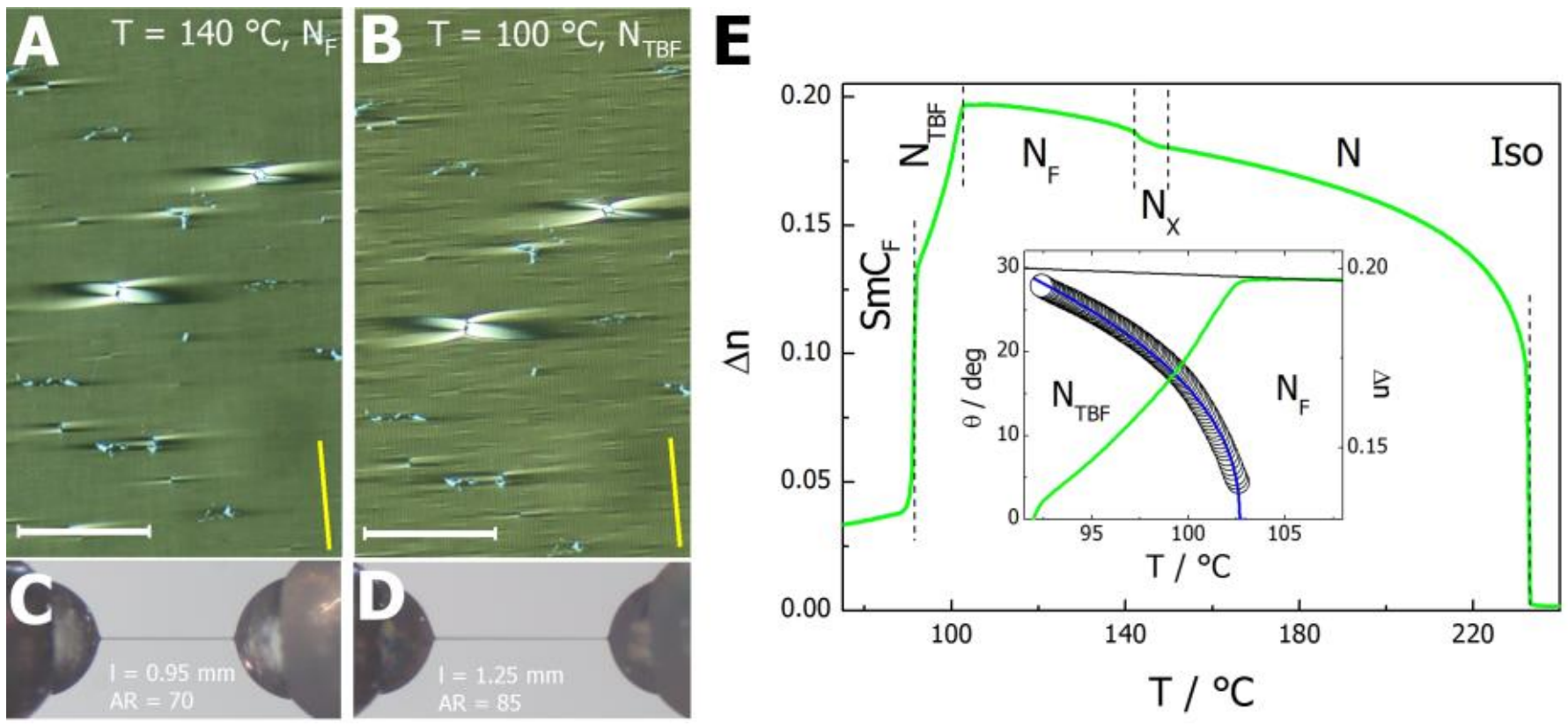

**Fig. 2**. **Optical studies.** Optical textures observed between crossed polarizers (at horizontal and vertical directions) for 1.5-μm-thick cell with parallel rubbing on both surfaces, the rubbing direction (indicated by yellow lines) was slightly inclined from vertical. Scale bars correspond to 100 μm. Images taken in ferroelectric nematic $N_F$ phase (**A**) and heliconical ferroelectric nematic phase, $N_{TBF}$ (**B**). Faint vertical stripes visible in **B** are oriented along the rubbing direction, visible defects are anchored at the plastic beads used as cell spacers. Fibers drawn in $N_F$ (**C**) and $N_{TBF}$ (**D**) phases between two glass capillaries, images were taken just before fiber rupture. AR stands for aspect ratio of the fiber. **E** Temperature dependence of optical birefringence, Δn (green line) measured for green light, λ=532 nm; note that Δn values measured in the smectic phase are strongly affected by the lack of sample alignment in this phase and resulting strong light scattering. In the inset, conical tilt angle (θ) in the heliconical structure of the $N_{TBF}$ phase (circles) calculated from decrease of Δn from its extrapolated temperature dependence (black line), the blue line is a guide for the eye.

## Optical birefringence

Further optical studies show that the phase transitions N-$N_x$-$N_F$ are accompanied by a slight increase in optical birefringence, and thus in the orientational order of the molecules (Fig. 2E). Below the $N_F$-$N_{TBF}$ phase transition, a gradual decrease in birefringence is observed. Together with the uniform texture indicating a uniaxial structure of the phase, this points to tilting and an azimuthal averaging of the molecular orientations by the formation of a heliconical structure, similar to that found in the $N_{TB}$ phase. Assuming formation of the heliconical structure, the conical tilt angle can be estimated from the decrease in birefringence (*23*). The tilt angle develops continuously and reaches 25 degrees at 10 K below the phase transition. The heliconical structure of the $N_{TBF}$ phase also explains the additional stabilization of the fibers against rupture compared to the fibers drawn in the $N_F$ phase.

## Light diffraction and selective reflection

Light diffraction experiments were carried out to obtain information about the pitch of the helix. In the 1.5-μm-thick cell with parallel rubbing, diffraction signals were observed in the direction perpendicular to the rubbing (Fig. 3A); their positions are nearly temperature-independent and correspond to the periodicity of the stripe texture observed in the microscopic studies (Fig. 2B). Moreover, additional diffraction signals were found at higher angles, which correspond to much lower periodicities and are aligned along the rubbing direction. The position of these additional diffraction signals was strongly temperature dependent; a few K below the $N_F$-$N_{TBF}$ phase transition the related periodicity was ~ 500–600 nm, and it was attributed to the helical pitch length. To avoid interaction with surfaces, the diffraction experiment was also performed on thicker cells without surfactant layers, on LC droplets deposited on glass and on freely suspended LC films. In all these samples ('powder' samples), a ring pattern with nearly the same diffraction intensity across the azimuthal angle was found (Fig. 3B). It changed its position with temperature (Fig. S9), in the range of angles corresponding to the periodicities of 550-1200 nm with increasing temperature (Fig. 3C). This proves that the helix unwinds critically when approaching the phase transition to the $N_F$ phase, and the changes were reversible upon heating and cooling.

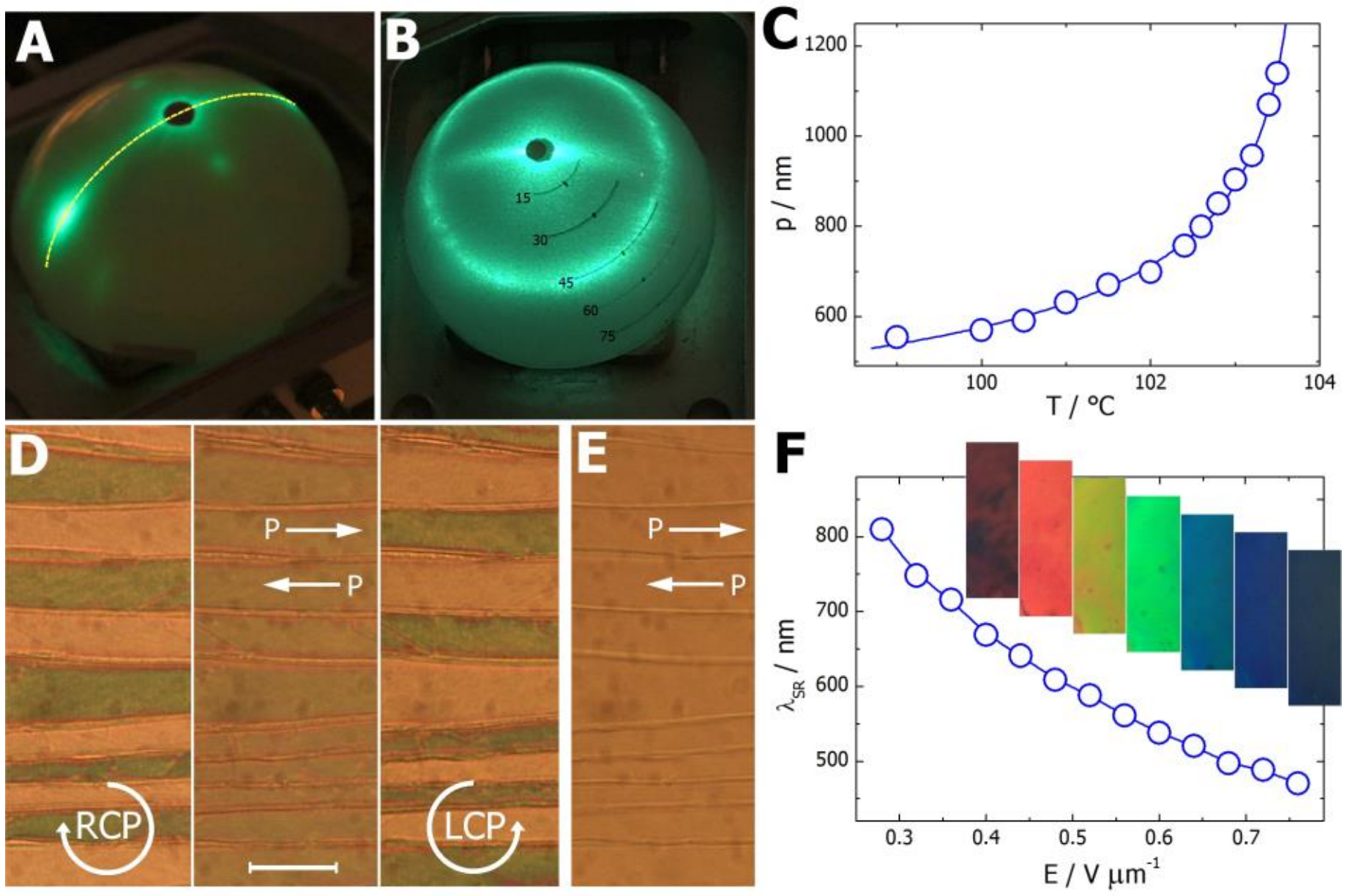

**Fig. 3**. **Light diffraction and selective reflection. AB** Diffraction patterns recorded in the $N_{TBF}$ phase, ~3 K below the $N_F$-$N_{TBF}$ phase transition, sample placed on a heating stage was illuminated from the bottom with a green (530 nm) laser, and diffracted light was projected onto spherical screen; **A** sample prepared in a 1.5-μm-thick cell with planar anchoring and parallel rubbing on both surfaces, yellow dashed line shows projection of the rubbing direction on the spherical screen, **B** sample prepared as one-free-surface droplet on untreated glass. **C** Temperature dependence of helical pitch (p) determined from diffraction experiments. **D** Images taken in $N_{TBF}$ phase (T=103 °C), a 10-μm-thick cell (with no alignment layers) was illuminated with white light (right-handed circularly polarized, unpolarized and left-handed circularly polarized for left, middle and right panels, respectively), greenish colors result from selective reflection (SR) of light. Only one type of domain shows SR phenomenon when illuminated with circularly polarized light. In $N_{TBF}$ the helix is tilted with respect to the surface as indicated by non-zero birefringence of the sample observed between crossed polarizers. Scale bar corresponds to 50 μm. **E** Image of the same sample area taken in $N_F$ phase (T=110 °C). It should be noted that visible domains with opposing directions of electric polarization (arrows) are not affected by the $N_{TBF}$ - $N_F$ phase transition. **F** Changes of selective light reflection wavelength, $\lambda_{SR}$, in the $N_{TBF}$ phase under application of dc electric field, E; in the inset: textures showing selective reflection colors visible for different values of applied voltage.

The helical structure of the $N_{TBF}$ phase gives rise to the selective reflection of light when the condition $\lambda_{SR} = n$p is fulfilled, where $n$ is an average refractive index and p is the helical pitch length. Scanning the transmission of light in a micrometer-size area, it was found that the center of the selective reflection band shifts from 700 to 440 nm with decreasing temperature (Fig. S10). The selective reflection is blue-shifted compared to the periodicities observed in the diffraction experiment because the helices are inclined to the direction of light propagation, as indicated by a low birefringence of the sample when observed between crossed polarizers. As expected, the selective reflection is sensitive to the handedness of circularly polarized light. When small uniform areas of a few micrometers are examined, the intensity of transmitted light decreases for only one of the circular polarizations, depending on the selected location in the sample (Fig. S10). When viewing larger areas illuminated with circularly polarized light, the domains that selectively reflect the light and those that do not are clearly distinguishable (Fig. 3D). Interestingly, the domain boundaries remain intact even after heating the sample into the $N_F$ phase (Fig. 3E). With the reasonable assumption that the direction of the electric polarization in the domains is maintained during the $N_{TBF}$-$N_F$ phase transition, one may speculate that the sense of the helix in the $N_{TBF}$ phase is related to the direction of the polarization.

The microscopic observations in the $N_{TBF}$ phase were also carried out under an electric field. The weak electric field reorients the molecules in the cell; when viewed between crossed polarizers the sample becomes homeotropic (with the optical axis along light propagation direction) and splits into micrometer-sized regions with optical activity of opposite sign (Fig. S11). A further increase of the electric field induces changes in the color of the selective light reflection in the sequence red, green, blue. The shift in the wavelength of the selective reflection is reversible when the electric field is increased or decreased (Fig. 3F, Fig. S12). Above the critical field (~500 mV $\mu m^{-1}$) the optical activity disappears. Although the shortening of the helix under the electric field might be surprising, given that the opposite effect is found in the cholesteric phase, similar changes have been reported for the twisted nematic phase made of dimers, in the materials showing the non-polar $N_{TB}$ phase (*24,25*). The application of an electric field shortens the helical pitch in the $N_{TBF}$ phase, and at the same time the cone angle becomes smaller.

### Second harmonic generation

Second harmonic generation (SHG) experiments were performed to confirm the non-centrosymmetric structure of the polar nematic phases. A strong SHG signal is found in the $N_F$ and $N_{TBF}$ phases (Fig. S13), however the signal decreases in the $N_{TBF}$ phase compared to the $N_F$ phase, which is consistent with the expected partial compensation of spontaneous polarization upon the formation of the heliconical structure. In the smectic phase, the signal is very weak, although non-zero, confirming that this phase is also polar. The second harmonic generation signal measured in $SmC_F$ phase is disturbed due to the strong scattering of the incident and generated light.

### Phase diagram

Finally, we investigated how the formation of the $N_{TBF}$ phase is influenced by changes in the molecular structure. The homologous series **MUT_JK10n** of mesogens was prepared, in which the terminal chain length, n, was modified, and a phase diagram was generated (Fig. 1B). Compared to previously studied four-ring DIO analogues (*26*), elongation of the mesogenic core increased the clearing temperature. What is somehow counterintuitive, is the decrease in the temperature of solid crystal melting, thus the temperature range of LC phase stability was significantly increased. The short homologs, n = 1 and n = 2, show a similar phase sequence of nematic phases as the **MUT_JK103** compound, but the $N_{TBF}$ phase becomes monotropic. For longer homologues, polar nematic phases are destabilized in favor of smectic phases: as expected molecules with longer terminal chains have a stronger propensity to segregate aromatic cores and alkyl chains in a lamellar structure. However, the tendency for polar ordering persists even for longer homologs. For **MUT_JK104**, with n=4, a sequence of N, SmA (non-polar), $SmA_F$, and $SmC_F$ phases was observed. Nevertheless, the temperature onset at which polar properties appear steeply decreases with elongation of the terminal chains, as dipole interactions decrease with the increasing distance between polar mesogenic cores. The stability of the twisted $N_{TBF}$ phase increases with the elongation of the terminal chain, indicating that interactions between mesogenic cores have to be somehow weakened to favor the twisted arrangement of dipoles over their co-linearity.

**Summary and conclusion**

We have observed a ferroelectric nematic phase, in which a sufficiently strong polar order leads to the spontaneous breaking of mirror symmetry. As a result, a heliconical structure is formed with the helical pitch in the NIR and visible light wavelength range. Although the structure of the $N_{TBF}$ phase is similar to the twist-bend nematic phase, the mechanism of its formation is different. While the non-polar $N_{TB}$ phase is driven by the decrease of bend elastic constant in the system, caused by the specific shape of the molecules, the $N_{TBF}$ phase is driven by polar interactions.

To understand the non-collinear arrangement of dipoles, it is necessary to consider the specific interactions between them. In solid ferroelectric or ferromagnetic materials, spatial variation of polarization or magnetization orientation is typically accommodated through the formation of discrete domains. Alternatively, more complex topologies may be formed, in which polarization/magnetization vectors change continuously over space without creating domain walls. In magnetic systems, in which exchange interactions between magnetic moments often dominate over crystal anisotropy, the spins can easily re-orient in the crystal lattice, and continuous structures, e.g. helices or vortexes, are common. They are usually attributed to the Dzyaloshinskii-Moriya interaction, permitted in crystals lacking inversion symmetry. This interaction yields homochiral spin topologies, with the handedness defined by crystal symmetry. Recent studies have shown that non-collinear magnetization structures, although rare, are also feasible in centrosymmetric crystals, e.g., due to long-range magnetodipolar interactions (*27*). In such systems, an additional degree of freedom is introduced through the formation of states with opposite chirality. Conversely, in solid ferroelectrics the space variation of electric polarization direction is related to the crystal lattice strain, thus complex polarization topologies are rare; electric skyrmions were found in thin membranes with specific boundary conditions (*28*), and helical ordering of electric dipoles was observed in quadruple perovskite crystals (*29*). Moreover, in ferroelectric systems there is no inherent mechanism for electric dipole interactions that would lift degeneracy of chirality sense, thus the overall pattern of electric dipoles exhibits a left- or right-handed rotation with equal probability (*30, 31*).

The nematic phase, with its lack of crystallographic structure, and thus low energy for reorientation of dipoles, resembles magnetic systems with their easily deformable magnetization. Therefore, one might expect the appearance of non-collinear, chiral arrangement of electric dipoles in polar nematic phases. However, unlike the magnetic DM interaction, electric dipolar interactions have to maintain degeneracy of chirality sense. For the material reported here, the pitch of the heliconical structure of the $N_{TBF}$ phase appears to be determined by electric interactions, yet the handedness is degenerate; both left- and right-handed helices are formed with equal probability. Nonetheless, our observations lay the groundwork for extrapolating concepts developed for magnetic systems and anticipating structures such as helices and skyrmions in polar soft systems as well.

## References

[1] M. Cestari, S. Diez-Berart, D. A. Dunmur, A. Ferrarini, M. R. de la Fuente, D. J. B Jackson, D. O. Lopez, G. R. Luckhurst, M. A. Perez-Jubindo, R. M. Richardson, J. Salud, B. A. Timimi, H. Zimmermann, Phase behavior and properties of the liquid-crystal dimer 1 ",7 "-bis(4-cyanobiphenyl-4 '- yl) heptane: A twist-bend nematic liquid crystal. *Phys Rev E* **84**, 031704 (2011)

[2] V. Borshch, Y. K. Kim, J. Xiang, M. Gao, A. Jakli, V. P. Panov, J. K. Vij, C. T. Imrie, M. G. Tamba, G. H, Mehl, O. D. Lavrentovich, Nematic twist-bend phase with nanoscale modulation of molecular orientation. *Nature Commun*. **4**, 2635 (2013).

[3] D. Chen, J. H. Porada, J. B. Hooper, A. Klittnick, Y. Shen, M. R. Tuchband, E. Korblova, D. Bedrov, D. M. Walba, M. A. Glaser, J. E. Maclennan, N. A. Clark, Chiral heliconical ground state of nanoscale pitch in a nematic liquid crystal of achiral molecular dimers. *Proc. Natl. Acad. Sci. USA* **110**, 15931-6 (2013).

[4] H. Nishikawa, K. Shiroshita, H. Higuchi, Y. Okumura, Y. Haseba, S. I. Yamamoto, K. Sago, H. Kikuchi, A Fluid Liquid-Crystal Material with Highly Polar Order. *Adv Mater* **29**, 1702354 (2017).

[5] R. J. Mandle, S. J. Cowling, J. W. Goodby, A nematic to nematic transformation exhibited by a rod-like liquid crystal. *Phys. Chem. Chem. Phys.* **19**, 11429-11435 (2017).

[6] A. Mertelj, L. Cmok, N. Sebastián, R. J. Mandle, R. R. Parker, A. C. Whitwood, J. W. Goodby, M. Čopič, Splay Nematic Phase. *Phys Rev X* **8**, 041025 (2018).

[7] H. Takezoe, E. Gorecka, M. Čepič, Antiferroelectric liquid crystals: Interplay of simplicity and complexity. *Rev. Mod. Phys.* **82**, 897-937 (2010).

[8] H. Takezoe, Y. Takanishi, Bent-core liquid crystals: their mysterious and attractive world. *Jpn. J. Appl. Phys*. **45**, 597 (2006).

[9] D. Pociecha, R. Walker, E. Cruickshank, J. Szydlowska, P. Rybak, A. Makal, J. Matraszek, J. M. Wolska, J. M. D. Storey, C. T. Imrie, E. Gorecka, Intrinsically chiral ferronematic liquid crystals: an inversion of the helical twist sense at the chiral nematic – chiral ferronematic phase transition. *J. Mol. Liq*., **361**, 119532 (2022).

[10] H. Nishikawa, F. Araoka, A New Class of Chiral Nematic Phase with Helical Polar Order. *Adv. Mat.* **33**, 2170270 (2021)

[11] X. Zhao, J. Zhou, J. Li, J. Kougo, Z. Wan, M. A.-O. Huang, S. A.-O. Aya, Spontaneous helielectric nematic liquid crystals: Electric analog to helimagnets. *Proc. Natl. Acad. Sci. USA*, **118**, e2111101118 (2021).

[12] C. Feng, R. Saha, E. Korblova, D. Walba, S. N. Sprunt, A. Jákli, Electrically Tunable Reflection Color of Chiral Ferroelectric Nematic Liquid Crystals. *Adv. Opt. Mat.* **9**, 2101230 (2021).

[13] T. Moriya, Anisotropic superexchange interaction and weak ferromagnetism. *Phys. Rev.* **120**, 91-98 (1960).

[14] A. Fert, N. Reyren, V. Cros, Magnetic skyrmions: advances in physics and potential applications. *Nat. Rev. Mater*. **2**, 17031 (2017).

[15] X. Chen, V. Martinez, E. Korblova, G. Freychet, M. Zhernenkov, M. A. Glaser, C. Wang, C. Zhu, L. Radzihovsky, J. E. Maclennan, D. M. Walba, N. A. Clark, The smectic ZA phase:

Antiferroelectric smectic order as a prelude to the ferroelectric nematic. *Proc. Natl. Acad. Sci. USA*, *120*, e2217150120 (2022).

[16] E. Cruickshank, P. Rybak, M. M. Majewska, S. Ramsay, C. Wang, C. Zhu, R. Walker, J. M. D. Storey, C. T. Imrie, E. Górecka, D. Pociecha, To Be or Not To Be Polar: The Ferroelectric and Antiferroelectric Nematic Phases. *ACS Omega* **8**, 36562-36568 (2023).

[17] R. Saha, P. Nepal, C. Feng, M. S. Hossain, M. Fukuto, R. Li, J. T. Gleeson, S. Sprunt, R. J. Twieg, A. Jákli, Multiple ferroelectric nematic phases of a highly polar liquid crystal compound, *Liq. Cryst.* **49**, 1784-1796 (2022).

[18] P. Kumari, B.Basnet, H. Wang, O. D. Lavrentovich, Ferroelectric nematic liquids with conics. *Nat. Commun.* **14**, 748 (2023).

[19] X. Chen, E. Korblova, M. A. Glaser, J. E. Maclennan, D. M. Walba, N. A. Clark, Polar in-plane surface orientation of a ferroelectric nematic liquid crystal: Polar monodomains and twisted state electro-optics. *Proc. Natl. Acad. Sci. USA* **118**, e2104092118 (2021).

[20] A. Jarosik, H. Nádasi, M. Schwidder, A. Manabe, M. Bremer, M. Klasen-Memmer, A. Eremin, Fluid fibres in true 3D ferroelectric liquids. *Proc. Natl. Acad. Sci. USA*, **121**, e2313629121 (2024)

[21] M. T. Máthé, K. Perera, Á. Buka, P. Salamon, A. Jákli, Fluid ferroelectric filaments, *Adv. Sci.* 2305950 (2023).

[22] D. H. Van Winkle, N. A. Clark, Freely Suspended Strands of Tilted Columnar Liquid Crystal Phases: One- Dimensional Nematics with Orientational Jumps. *Phys. Rev. Lett*. **48**, 1407–1411 (1982)

[23] C. Meyer, G. R. Luckhurst, I. Dozov, The temperature dependence of the heliconical tilt angle in the twist-bend nematic phase of the odd dimer CB7CB. *J. Mater. Chem. C* **3**, 318-328, (2015).

[24] J. Xiang, Y. Li , Q. Li, D. A. Paterson, J. M. Storey, C. T. Imrie, O. D. Lavrentovich, Electrically tunable selective reflection of light from ultraviolet to visible and infrared by heliconical cholesterics. *Adv. Mater.* **27**, 3014-8 (2015)

[25] C. Yuan, W. Huang, Z. Zheng, B. Liu, H. K. Bisoyi, Y. Li, D. Shen, Y. Lu, Q. Li, Stimulated transformation of soft helix among helicoidal, heliconical and their inverse helices. *Sci. Adv.* **5**, aax9501 (2019).

[26] J. Li, H. Nishikawa, J. Kougo, J. Zhou, S. Dai, W. Tang, X. Zhao, Y. Hisai, M. Huang, S. Aya, Development of ferroelectric nematic fluids with giant-ε dielectricity and nonlinear optical properties. *Sci. Adv.* **7**, eabf5047 (2021).

[27] X. Yu, M. Mostovoy, Y. Tokunaga, W. Zhangc, K. Kimoto, Y. Matsui, Y. Kaneko, N. Nagaosa, Y. Tokura, Magnetic stripes and skyrmions with helicity reversals. *Proc. Natl. Acad. Sci. USA* **109**, 8856-8860 (2012).

[28] S. Das, Y. L. Tang, Z. Hong, M. A. P. Gonçalves, M. R. McCarter, C. Klewe, K. X. Nguyen, F. Gómez-Ortiz, P. Shafer, E. Arenholz, V. A. Stoica, S.-L. Hsu, B. Wang, C. Ophus, J. F. Liu, C. T. Nelson, S. Saremi, B. Prasad, A. B. Mei, D. G. Schlom, J. Íñiguez, P. García-Fernández, D. A. Muller, L. Q. Chen, J. Junquera, L. W. Martin, R. Ramesh, Observation of room-temperature polar skyrmions. *Nature* **568**, 368–372 (2019).

[29] H. J. Zhao, P. Chen, S. Prosandeev, S. Artyukhin, L. Bellaiche, Dzyaloshinskii–Moriya-like interaction in ferroelectrics and antiferroelectrics. *Nat. Mater.* **20**, 341–345 (2021).

[30] Y.-T. Shao, S. Das, Z. Hong, R. Xu, S. Chandrika, F. Gómez-Ortiz, P. García-Fernández, L.-Q. Chen, H. Y. Hwang, J. Junquera, L. W. Martin, R. Ramesh, D. A. Muller, Emergent chirality in a polar meron to skyrmion phase transition *Nat. Commun.* **14**, 1355 (2023).

[31] S. Das, Z. Hong, M. McCarter, P. Shafer, Y.-T. Shao, D. A. Muller, L. W. Martin, R. Ramesh, A new era in ferroelectrics. *APL Mater*. **8**, 120902 (2020)

[32] J. C. Kemp, Polarized light and its interaction with modulating devices, Hinds International, Inc., (1987).

[33] Gaussian 16, Revision C.01, M. J. Frisch, G. W. Trucks, H. B. Schlegel, G. E. Scuseria, M. A. Robb, J. R. Cheeseman, G. Scalmani, V. Barone, G. A. Petersson, H. Nakatsuji, X. Li, M. Caricato, A. V. Marenich, J. Bloino, B. G. Janesko, R. Gomperts, B. Mennucci, H. P. Hratchian, J. V. Ortiz, A. F. Izmaylov, J. L. Sonnenberg, D. Williams-Young, F. Ding, F. Lipparini, F. Egidi, J. Goings, B. Peng, A. Petrone, T. Henderson, D. Ranasinghe, V. G. Zakrzewski, J. Gao, N. Rega, G. Zheng, W. Liang, M. Hada, M. Ehara, K. Toyota, R. Fukuda, J. Hasegawa, M. Ishida, T. Nakajima, Y. Honda, O. Kitao, H. Nakai, T. Vreven, K. Throssell, J. A. Montgomery, Jr., J. E. Peralta, F. Ogliaro, M. J. Bearpark, J. J. Heyd, E. N. Brothers, K. N. Kudin, V. N. Staroverov, T. A. Keith, R. Kobayashi, J. Normand, K. Raghavachari, A. P. Rendell, J. C. Burant, S. S. Iyengar, J. Tomasi, M. Cossi, J. M. Millam, M. Klene, C. Adamo, R. Cammi, J. W. Ochterski, R. L. Martin, K. Morokuma, O. Farkas, J. B. Foresman, and D. J. Fox, Gaussian, Inc., Wallingford CT, 2016.

[34] A. D. Becke, Density-functional thermochemistry. III. The role of exact exchange. *J. Chem. Phys.* **98**, 5648–5652 (1993).

[35] C. Lee, W. Yang, R. G. Parr, Development of the Colle-Salvetti correlation-energy formula into a functional of the electron density. *Phys. Rev.* **37**, 785–789 (1988).

[36] S. Grimme, J. Antony, S. Ehrlich, H. Krieg, A consistent and accurate ab initio parametrization of density functional dispersion correction (DFT-D) for the 94 elements H-Pu. *J. Chem. Phys.* **132** (2010).

[37] E. Papajak, J. Zheng, X. Xu, H. R. Leverentz, D. G. Truhlar, Perspectives on basis sets beautiful: seasonal plantings of diffuse basis functions. *J. Chem. Theory Comput.* **7**, 3027–3034 (2011).

[38] C. F. Macrae, I. Sovago, S. J. Cottrell, P. T. A. Galek, P. McCabe, E. Pidcock, M. Platings, G. P. Shields, J. S. Stevens, M. Towler and P. A. Wood, Mercury 4.0: from visualization to analysis, design and prediction. *J. Appl. Cryst*. **53**, 226-235, 2020

[39] L. Tian, F. Chen, Multiwfn: A multifunctional wavefunction analyzer. *J. Comp. Chem*. **33**, 580–592 (2011).

[40] J. Karcz, N. Rychłowicz, M. Czarnecka, A. Kocot, J. Herman, P. Kula, Enantiotropic ferroelectric nematic phase in a single compound. *Chem. Commun.,* **59**, 14807-14810 (2023).

[41] X. Chen, E. Korblova, D. Dong, X. Wei, R. Shao, L. Radzihovsky, M. A. Glaser, J. E. Maclennan, D. Bedrov, D. M. Walba, N. A. Clark, First-principles experimental demonstration of ferroelectricity in a thermotropic nematic liquid crystal: Polar domains and striking electro-optics, *Proc. Natl. Acad. Sci. USA*, **117**, 14021–14031 (2020)

**Acknowledgments**

Authors thank to prof. N. Vaupotič for discussion and comments. We gratefully acknowledge Polish high-performance computing infrastructure PLGrid (HPC Centers: ACK Cyfronet AGH, WCSS) for providing computer facilities and support within computational grant no. PLG/2023/016670.

**Funding:**

National Science Centre (Poland), grant no. 2021/43/B/ST5/00240 (DP)

Military University of Technology project no. UGB 22-801 (PK)

# Supplementary Materials

## Experimental methods

For *differential scanning calorimetry* (DSC) studies, a TA Q200 calorimeter was used, calibrated using indium and zinc standards. Heating and cooling rates were 5-20 K $min^{-1}$, samples were kept in a nitrogen atmosphere. The transition temperatures and associated thermal effects were extracted from the heating traces.

*X-ray diffraction* (XRD) studies in broad diffraction angle range were performed with Bruker GADDS system equipped with micro-focus type X-ray tube with Cu anode, and Vantec 2000 area detector. Samples were prepared in the form of small drops placed on a heated surface, their temperature was controlled with a modified Linkam heating stage. For small angle diffraction experiments Bruker Nanostar system was used (micro-focus type X-ray tube with Cu anode, MRI TCPU-H heating stage, Vantec 2000 area detector). Samples were prepared in thin-walled glass capillaries, with 1.5 mm diameter.

*Optical* textures of LC phases were studied using a Zeiss Axio Imager A2m polarized light microscope, equipped with a Linkam TMS 92 heating stage. Samples were prepared in commercial cells (AWAT) of various thicknesses (1.5–20 μm) with ITO electrodes and surfactant layers for planar or homeotropic alignment, in the case of planar cells either parallel or antiparallel rubbing on both surfaces was applied.

*Optical* b*irefringence* was measured with a setup based on a photoelastic modulator (PEM-90, Hinds) working at the base frequency $f$=50 kHz. As a light source, a halogen lamp (Hamamatsu LC8) equipped with a narrow bandpass filter (532±3 nm) was used. Samples were prepared in glass cells with a thickness of 1.5 μm, having surfactant layers for planar anchoring condition, and parallel rubbing assuring uniform alignment of the optical axis in nematic phases. The sample and PEM were placed between crossed linear polarizers, with axes rotated ±45 deg with respect to the PEM axis, and the intensity of the light transmitted through this set-up was measured with a photodiode (FLC Electronics PIN-20). The registered signal was de-convoluted with a lock-in amplifier (EG&G 7265) into $1f$ and $2f$ components to yield a retardation induced by the sample, (*32*). Based on the measured optical birefringence the conical tilt angle ($\theta$) in the twist-bend ferroelectric nematic phase ($N_{TBF}$) was deduced from the decrease of the $\Delta n$ with respect to the values measured in the ferroelectric nematic ($N_F$) phase, according to the relation: $\Delta n_{N_{TBF}} = \Delta n_{NF}(3\cos^2\theta - 1)/2$ (*23*). The birefringence of the ferroelectric nematic phase was extrapolated to the lower temperature range by assuming a power law temperature dependence: $\Delta n_{NF} = \Delta n_0 (T_c - T)^\gamma$, where $\Delta n_0$, $T_c$, and $\gamma$ are the fitting parameters.

*Optical diffraction* studies were performed for the samples placed on a heating stage and illuminated from below with green (520 nm) laser light. The diffraction pattern was recorded on the half-sphere screen placed above the sample, the angular position of the observed diffraction signal allowed for calculation of the related periodicity of the stripe pattern in the cell, and helical pitch length.

*Selective light reflection* studies were carried out for material placed in the glass cell with ITO electrodes, either having surfactant layer assuring planar/homeotropic anchoring prior to application of electric field or without alignment layers. Light transmission/reflection was monitored with CRAIC 20/20 PV microspectrophotometer, equipped with Linkam heating stage. The measurements were performed with ~20 micron thick cells with homeotropic anchoring. The tested sample area was confined to 50 microns. Experiments were also performed with a fiber-coupled spectrometer (Filmetrics F20-UV) mounted to the Zeiss Axio Imager A2m microscope, which allowed for use of circularly polarized light.

*Fiber drawing:* The fibers were drawn between two capillaries sealed at the ends, immersed before experiment in the LC material to provide sufficiently large material reservoir. To ensure temperature control, the capillaries were placed in a small box having glass walls with transparent ITO heaters. The aspect ratio, at which the fiber rupture occurred, was obtained by analyzing the selected frames from a movie recorded during the process of fiber thinning.

*Electric polarization measurements* were performed using cells with ITO electrodes and no surfactant layers. The spontaneous polarization was calculated by analyzing the current flow through a resistor (500 Ω) connected in series with the cell under testing, upon application of triangular-wave voltage. Siglent SDG2042X arbitrary waveform generator, FLC A200 amplifier and Siglent SDS2000X Plus oscilloscope were used.

*Dielectric spectroscopy*: The complex dielectric permittivity was measured in the 1 Hz–10 MHz frequency range using a Solartron 1260 impedance analyzer. The material was placed in 3-μm-thick glass cell with ITO electrodes (without the polymer alignment layers to avoid the influence of the high capacitance of a thin polymer layer). The amplitude of the applied ac voltage, 20 mV, was low enough to avoid Fréedericksz transition in nematic phases.

*Second Harmonic Generation*: The SHG response was investigated using a setup based on a solid-state IR laser EKSPLA NL202, λ=1064 nm. A collimated (~1 mm $1/e^2$ ) series of 9 ns laser pulses at a 10 Hz repetition rate and <2 mJ pulse energy were applied. The pulse energy was adjusted to avoid material decomposition. The IR beam was incident onto an LC cell of thickness 10 μm. An IR pass filter was placed at the entrance to the sample and a green pass filter at the exit of the sample. The emitted SHG radiation was detected using a photon counting head (Hamamatsu H7421) with a power supply unit (C8137). The SHG signal intensity was monitored with an oscilloscope (Agilent Technologies DSO6034A) using a custom-written Python script.

*Molecular Modeling*: The geometric parameters were calculated with quantum mechanical calculations using the Gaussian 16 (Revision C.01) software package (*33*) on the PLGrid ASK Cyfronet Ares cluster. Geometry of the molecule was calculated with B3LYP (*34, 35*) hybrid DFT functional with Grimme's D3 dispersion correction (*36*) and may-cc-pVTZ basis set (*37*). Frequency calculations were performed to confirm that the obtained geometry corresponds to the true minimum. Single-point calculation on the same level of theory was performed to obtain the dipole moment. Output files were rendered using Mercury (*38*). The size of the molecule was determined using the Multiwfn software (*39*) – the molecule was rotated to make its principal axes parallel to the three Cartesian axes and boundary atoms were determined. For the new coordinate system, the angle between the direction of the dipole moment and molecular axis (X axis) was calculated.

*Nuclear Magnetic Resonance*: the proton ($^1$H), carbon ($^{13}$C), and fluorine ($^{19}$F) NMR spectra in $CDCl_3$ or DMSO-*d6* were collected using a Bruker, model Avance III spectrometer (Bruker, Billerica, MA, USA).

## Organic synthesis

The molecular structure of the studied here material, 4'-(difluoro(3,4,5-trifluorophenoxy)methyl)-2,3',5'-trifluoro-[1,1'-biphenyl]-4-yl 2,6-difluoro-4-(5-propyl-1,3-dioxan-2-yl)benzoate, referred to later as **MUT_JK103**, synthesized by the Liquid Crystal group at the Military University of Technology, is based on DIO – one of the model ferronematogens (*4*). The synthetic route was divided into two parts (Fig. S1); in the first part, 3,5-difluorobenzaldehyde **(2)** was synthesized by formylation of the previously obtained Grignard reagent. Next, the 1,3-dioxane ring was introduced into the molecule **(3)**, which was later lithiated and carboxylated producing the corresponding carboxylic acid **(4)**. In the second part, in the Suzuki-Miyaura reaction the phenol **(7)** was obtained by reacting boronic ester **(6)**, obtained previously from bromo derivative **(5)** in a Miyaura borylation reaction, with the 4-bromo-3-fluorophenol. Finally, **MUT_JK103** compound was synthesized in a Steglich esterification reaction between **(4)** and **(7)** with the presence of DCC and DMAP. The final compound was purified using the recrystallization technique. The synthesis of **(2)**, **(3)**, **(4)** and **(6)** has been described previously (*40*). The other homologs of **MUT_JK10n** series were synthesized using analogous methods.

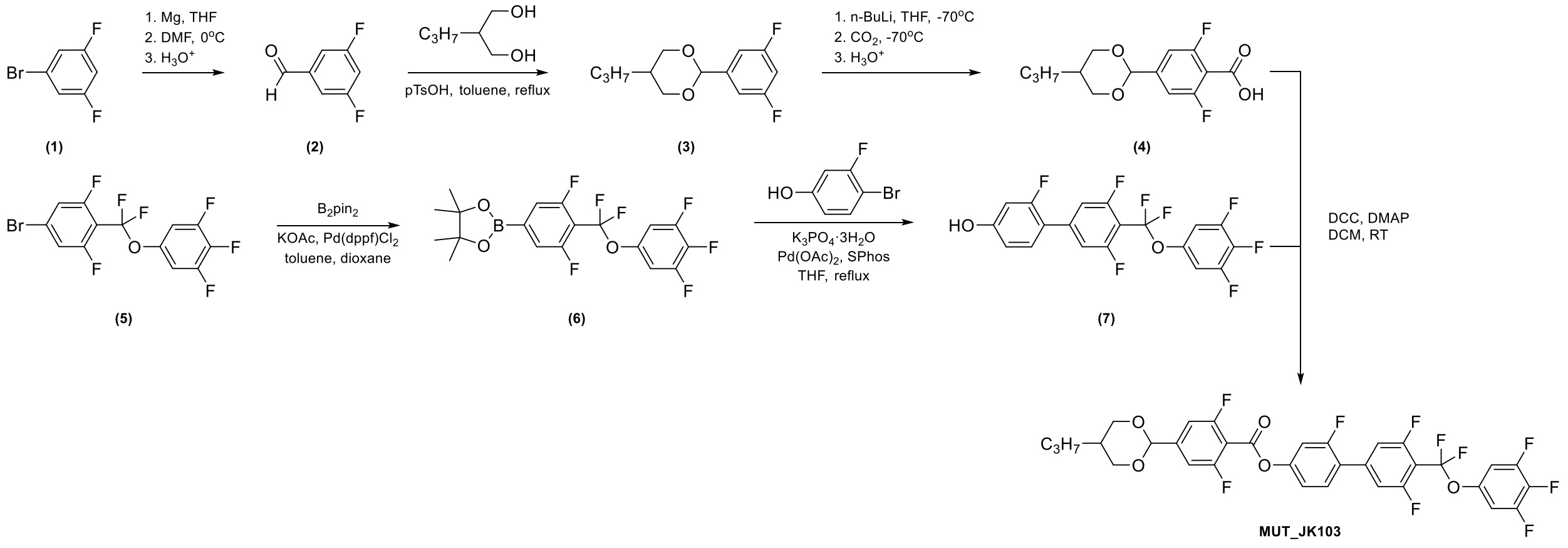

**Figure S1.** Synthesis of studied compound **MUT_JK103**

***4'-(difluoro(3,4,5-trifluorophenoxy)methyl)-2,3',5'-trifluoro-[1,1'-biphenyl]-4-ol (7)***

The mixture of 2-(4-(difluoro(3,4,5-trifluorophenoxy)methyl)-3,5-difluorophenyl)-4,4,5,5-tetramethyl-1,3,2-dioxaborolane **(6)** (5.71 g; 0.0131 mol), 4-bromo-3-fluorophenol (2.27 g; 0.0119 mol), potassium phosphate trihydrate (11.11 g; 0.0417 mol) in anhydrous THF (100 ml) was refluxed for 1h under $N_2$ atmosphere. Then $Pd(OAc)_2$ and SPhos were added and the mixture was stirred at reflux for 2h. Later, it was cooled down, acidified using 10% HCl solution and extracted with DCM. The organic layer was then dried over $MgSO_4$, and concentrated under vacuum. The residue was recrystallized from hexane to give an off-white solid.

Yield 4.27g (85.4%)

Purity 96.9% (GC-MS)

m.p. = 79°C

MS(EI) m/z: 420 (M+); 401; 273; 244; 224; 204; 175; 156; 136; 119

$^{1}$H NMR (500 MHz, DMSO-*d6*) δ: 10.41 (s, 1H, Ar-OH); 7.49 (t, J=9.15 Hz, 1H, Ar-H); 7.42 (d, J=11.60 Hz, 2H, Ar-H); 7.36 (dd, J=7.93, 6.10 Hz, 2H, Ar-H); 6.74 (dd, J=8.55, 2.14 Hz, 1H, Ar-H); 6.69 (dd, J=13.12, 2.44 Hz, 1H, Ar-H)

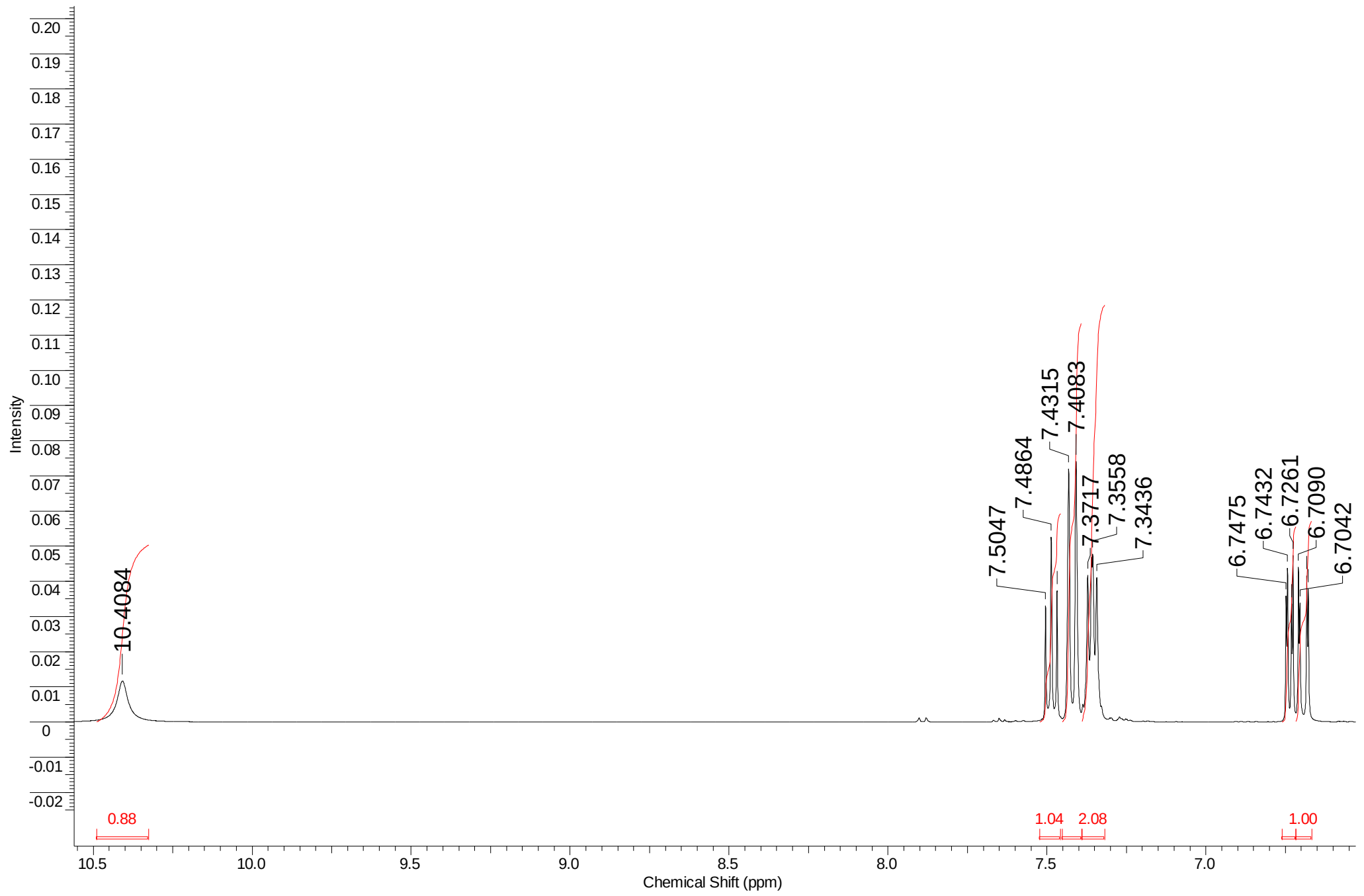

$^{13}$C NMR (125 MHz, DMSO-*d6*) δ: 161.42; 160.84 (d, *J*=11.81 Hz); 160.50 (d, *J*=6.36 Hz); 159.44; 158.48 (d, *J*=6.36 Hz); 150.71 (dq, *J*=248.76, 5.34, 5.22, 5.00 Hz); 144.63 (m); 142.52 (t, *J*=11.35 Hz); 138.26 (dt, *J*=247.96, 15.44 Hz); 131.69 (d, *J*=4.54 Hz); 120.45 (t, *J*=264.31 Hz); 115.49 (d, *J*=11.81 Hz); 112.96 (m); 108.25 (m); 106.90 (m); 103.70 (d, *J*=25.43 Hz)

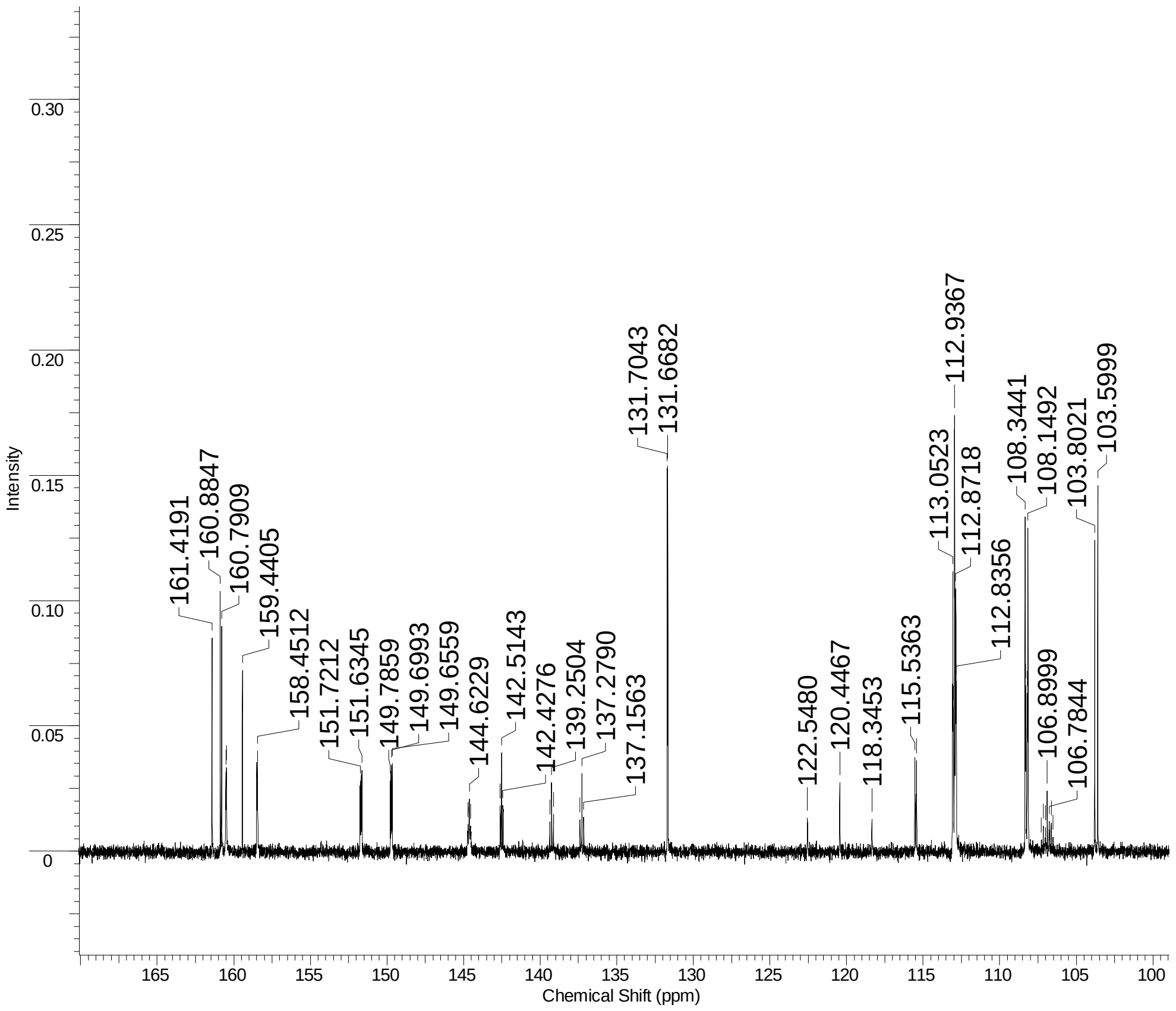

$^{19}F$ NMR (471 MHz, $CDCl_3$) δ: -61.69 (t, J=26.0 Hz, 2F); -111.13 (td, J=26.0, 10.4 Hz, 2F); -114.70 (t, J=10.4 Hz, 1F); -132.53 (dd, J=20.8, 8.7 Hz, 2F); -163.22 (tt, J=20.8, 5.2 Hz, 1F)

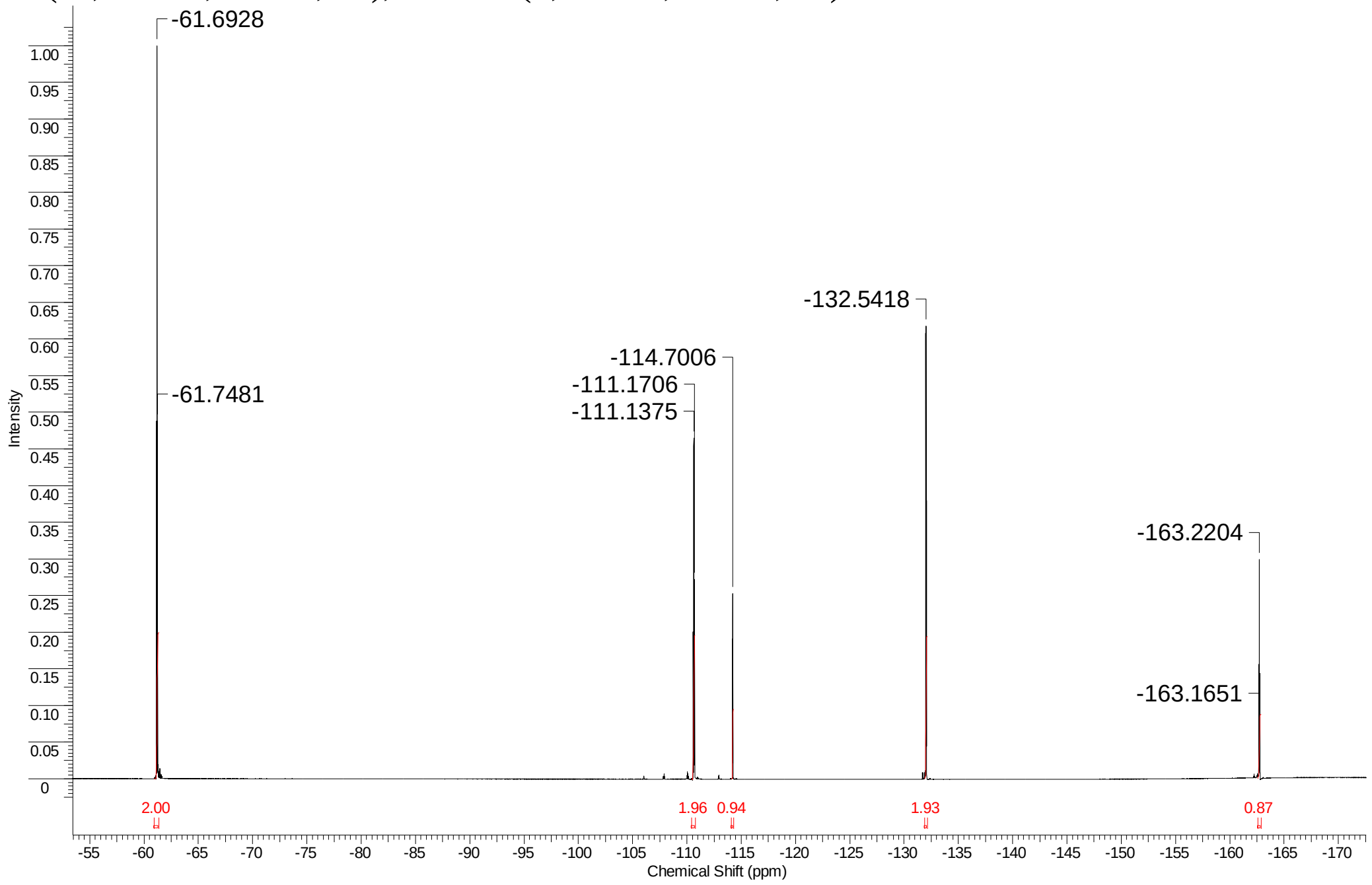

***4'-(difluoro(3,4,5-trifluorophenoxy)methyl)-2,3',5'-trifluoro-[1,1'-biphenyl]-4-yl 2,6-difluoro-4-(5-propyl-1,3-dioxan-2-yl)benzoate (MUT_JK103)***

To a stirred solution of 2,6-difluoro-4-(5-propyl-1,3-dioxan-2-yl)benzoic acid **(4)** (1.24 g; 0.00433 mol), 4'-(difluoro(3,4,5-trifluorophenoxy)methyl)-2,3',5'-trifluoro-[1,1'-biphenyl]-4-ol **(7)** (2 g; 0.00476 mol) and DCC (0.98 g; 0.00476 mol) in DCM (50ml), DMAP (0.1g) was added and the solution was stirred overnight at room temperature. The reaction mixture was filtered through silica pad and the filtrate was concentrated under vacuum. The residue was then recrystallized from ethanol/acetone mixture to give white solid.
Yield 2.1g (70.5%)

Purity 99.6% (99.6/0.2 trans/cis ratio) (HPLC-MS)

MS(EI) m/z: 687 (M+); 669; 590; 569; 541; 441; 419; 384; 269; 244; 220; 193; 169; 141

$^1$H NMR (500 MHz, $CDCl_3$) δ: 7.51 (t, J=8.70 Hz, 1 H); 7.23 (m, 6 H); 7.02 (dd, J=7.32, 6.10 Hz, 2 H); 5.43 (s, 1 H); 4.28 (dd, J=11.90, 4.58 Hz, 2 H); 3.56 (t, J=11.44 Hz, 2 H); 2.16 (m, 1 H); 1.37 (m, 2 H); 1.12 (m, 2 H); 0.96 (t, J=7.32 Hz, 3 H)

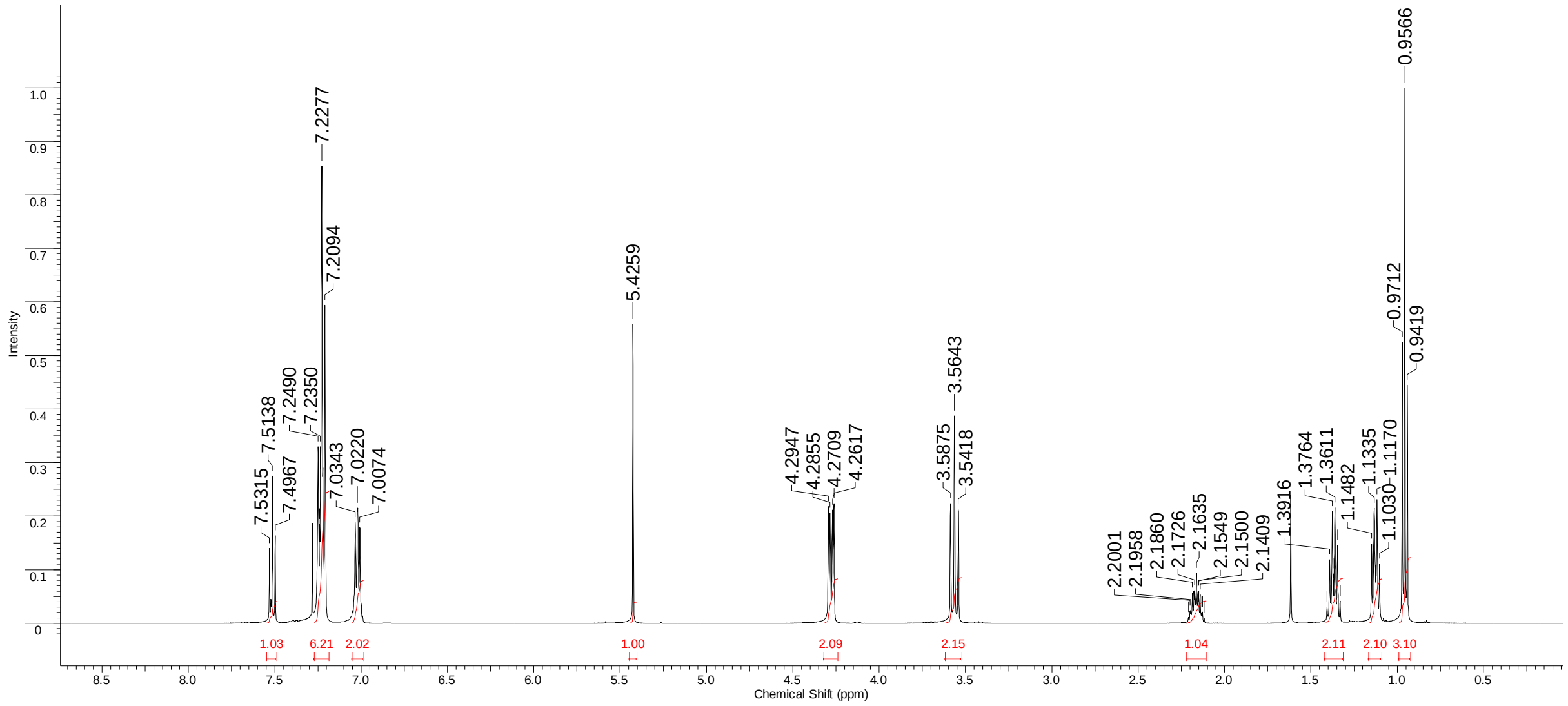

$^{13}C$ NMR (125 MHz, $CDCl_3$) δ: 161.98 (d, J=5.45 Hz); 160.98 (d, J=6.36 Hz); 160.52; 159.93 (d, J=5.45 Hz); 159.19; 158.93 (d, J=6.36 Hz); 158.52; 152.02 (q, J=10.90, 5.45 Hz); 151.66 (d, J=10.90 Hz); 150.03 (q, J=10.90, 5.45 Hz); 145.79 (t, J=9.99 Hz); 144.62 (m); 140.69 (t, J=10.90 Hz); 138.49 (dt, J=250.69, 15.44 Hz); 130.62 (d, J=3.63 Hz); 123.61 (d, J=12.72 Hz); 120.15 (t, J=266.13 Hz); 118.36 (d, J=3.63 Hz); 113.13 (dd, J=27.70, 3.18 Hz); 110.94; 110.73; 110.33 (m); 109.09 (m); 107.50 (m); 98.79 (t, J=2.27 Hz); 72.60; 33.92; 30.23; 19.55; 14.18

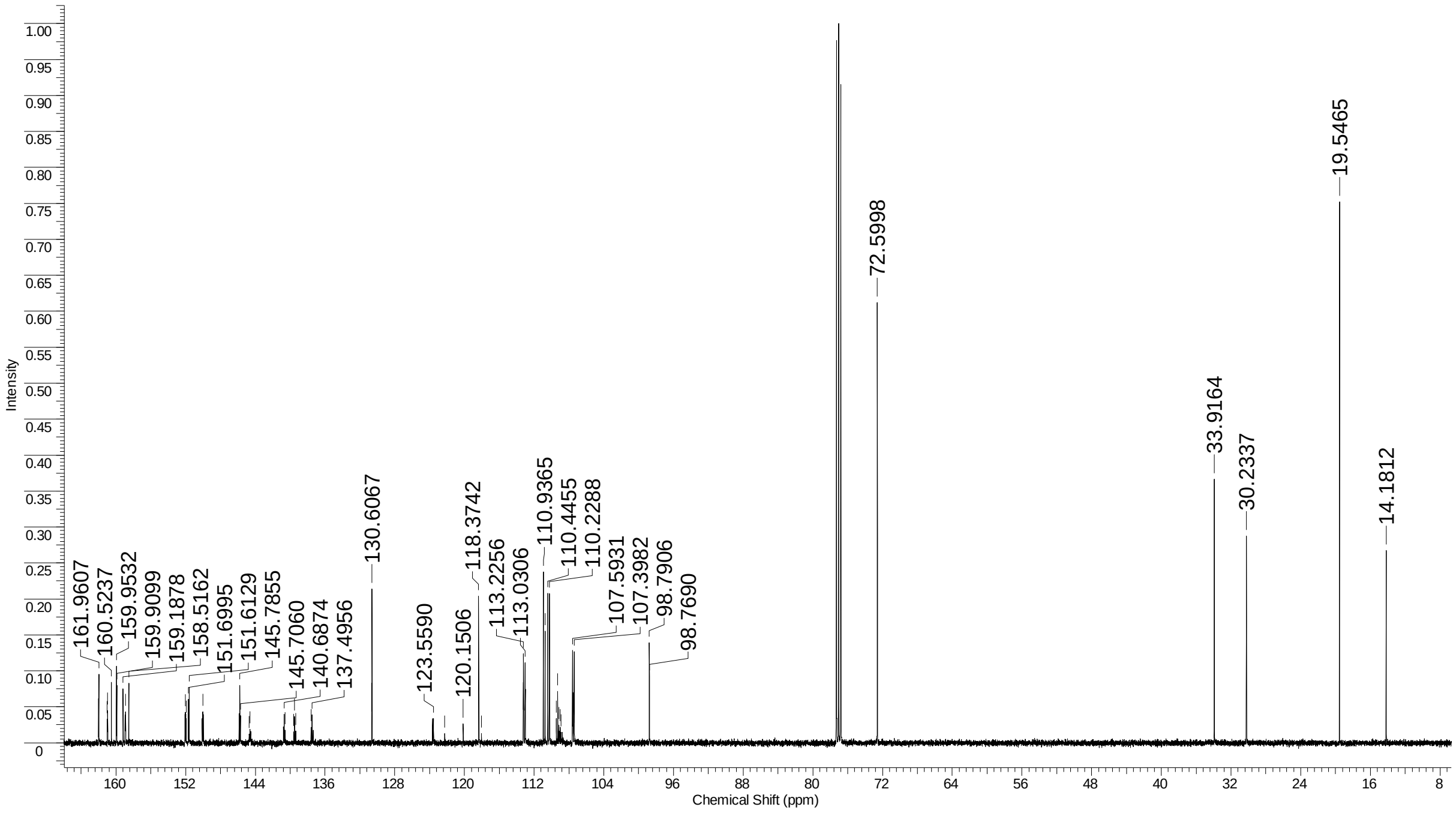

$^{19}F$ NMR (471 MHz, $CDCl_3$) δ: -61.84 (t, J=26.0 Hz, 2F); -108.36 (d, J=8.7 Hz, 2F); -110.37 (td, J=26.0, 10.4 Hz, 2F); -113.64 (t, J=10.4 Hz, 1F); -132.43 (dd, J=20.8, 7.0 Hz, 2F); -163.08 (tt, J=20.8, 5.2 Hz, 1F)

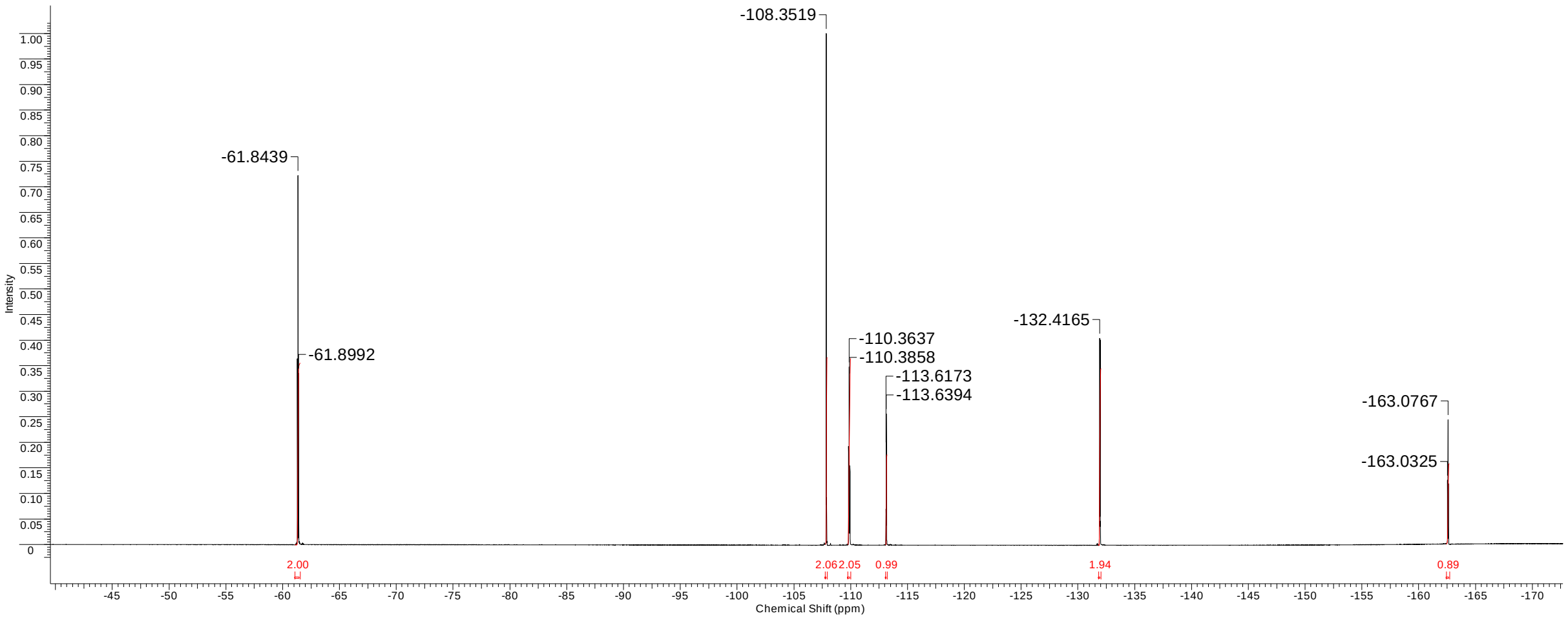

***4'-(difluoro(3,4,5-trifluorophenoxy)methyl)-2,3',5'-trifluoro-[1,1'-biphenyl]-4-yl 2,6-difluoro-4-(5-methyl-1,3-dioxan-2-yl)benzoate (MUT_JK101)***

Yield 2.44g (76.3%)

Purity 99.5% (99.5/0.5 trans/cis ratio) (HPLC-MS)

MS(EI) m/z: 660 (M+); 642; 587; 569; 513; 441; 419; 384; 313; 272; 241; 220; 196; 169; 141

$^1$H NMR (500 MHz, $CDCl_3$) δ: 7.51 (t, *J*=8.55 Hz, 1 H); 7.23 (m, 6 H); 7.02 (m, 2 H); 5.43 (s, 1 H); 4.23 (dd, *J*=11.60, 4.88 Hz, 2 H); 3.54 (t, *J*=11.44 Hz, 2 H); 2.25 (m, 1 H); 0.82 (d, *J*=6.71 Hz, 3 H)

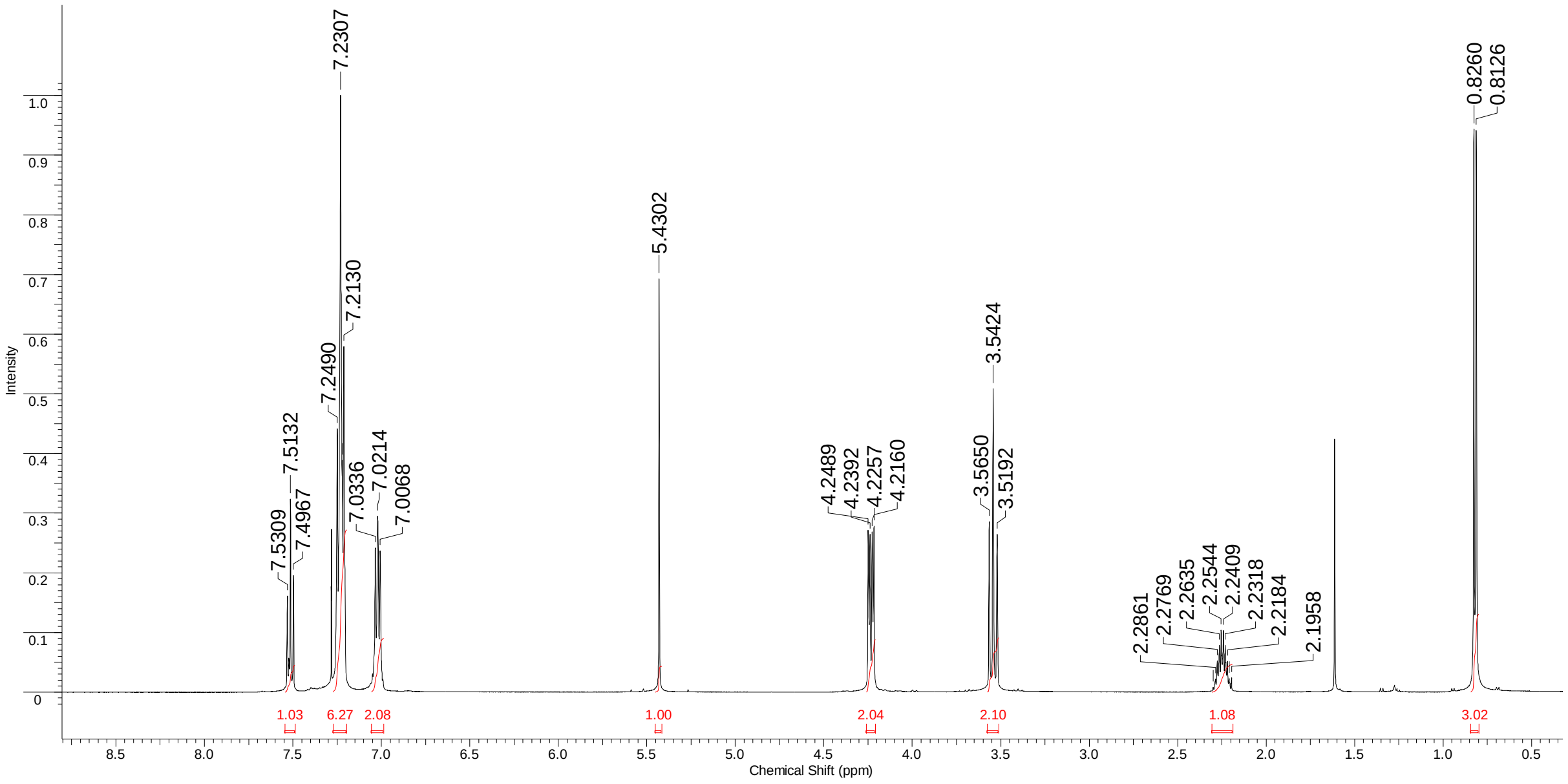

$^{13}C$ NMR (125 MHz, $CDCl_3$) δ: 161.99 (d, *J*=5.45 Hz); 160.98 (d, J=6.36 Hz); 160.52; 159.93 (d, *J*=5.45 Hz); 159.21; 158.93 (d, *J*=6.36 Hz); 158.52; 152.02 (q, *J*=10.90, 5.45 Hz); 151.64 (d, *J*=11.81 Hz); 150.03 (q, *J*=10.90, 5.45 Hz); 145.70 (t, *J*=9.99 Hz); 144.60 (m); 140.68 (t, J=9.99 Hz); 138.49 (dt, *J*=250.69, 15.44 Hz); 130.64 (d, *J*=3.63 Hz); 123.65 (d, *J*=12.72 Hz); 120.14 (t, J=266.13 Hz); 118.37 (d, *J*=3.63 Hz); 113.15 (dd, J=27.08, 3.63 Hz); 110.96; 110.75; 110.36 (m); 109.25 (m); 107.50 (m); 98.57 (t, J=2.73 Hz); 73.62; 29.29; 12.25

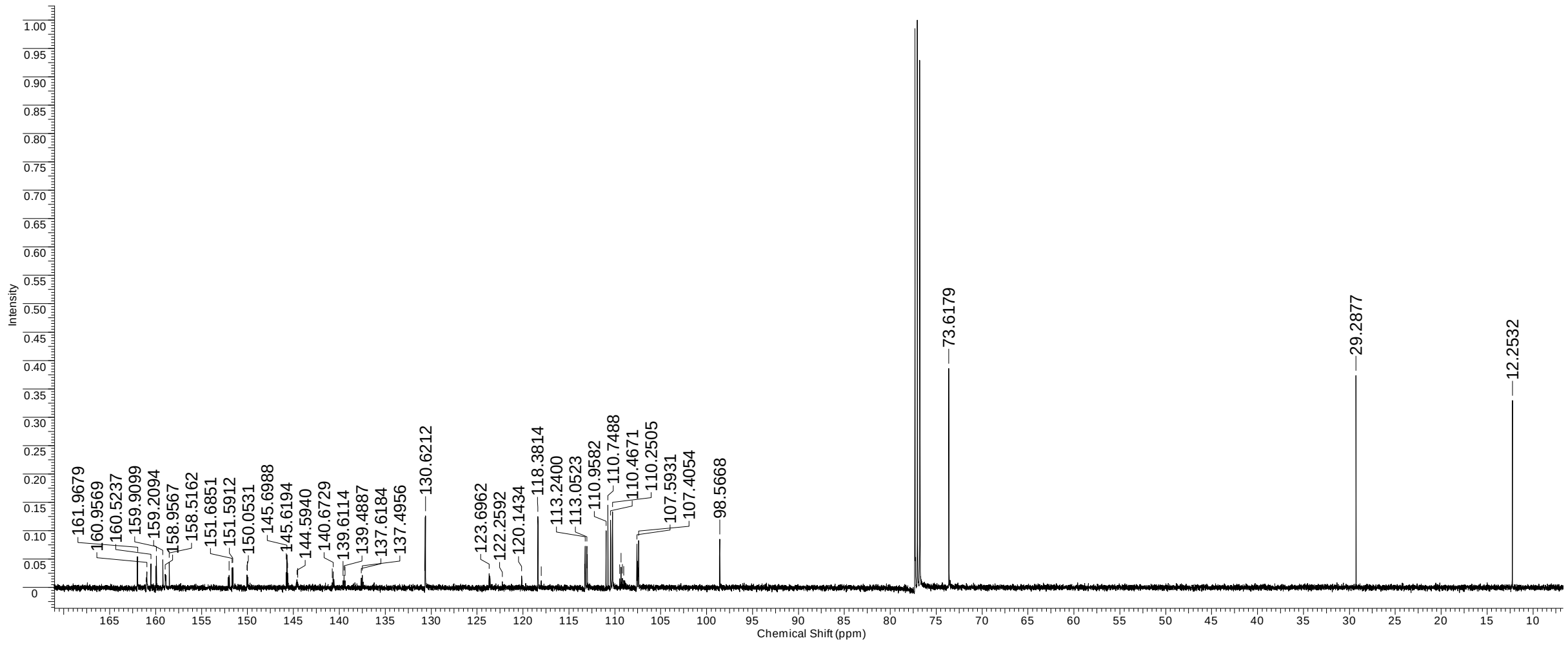

$^{19}F$ NMR (471 MHz, $CDCl_3$) δ: -61.84 (t, J=26.0 Hz, 2 F); -108.36 (d, J=8.7 Hz, 2 F); -110.37 (td, J=26.0, 10.4 Hz, 2 F); -113.64 (t, J=10.4 Hz,1 F); -132.43 (dd, J=20.8, 8.7 Hz, 2 F); -163.07 (tt, J=20.8, 5.2 Hz,1 F)

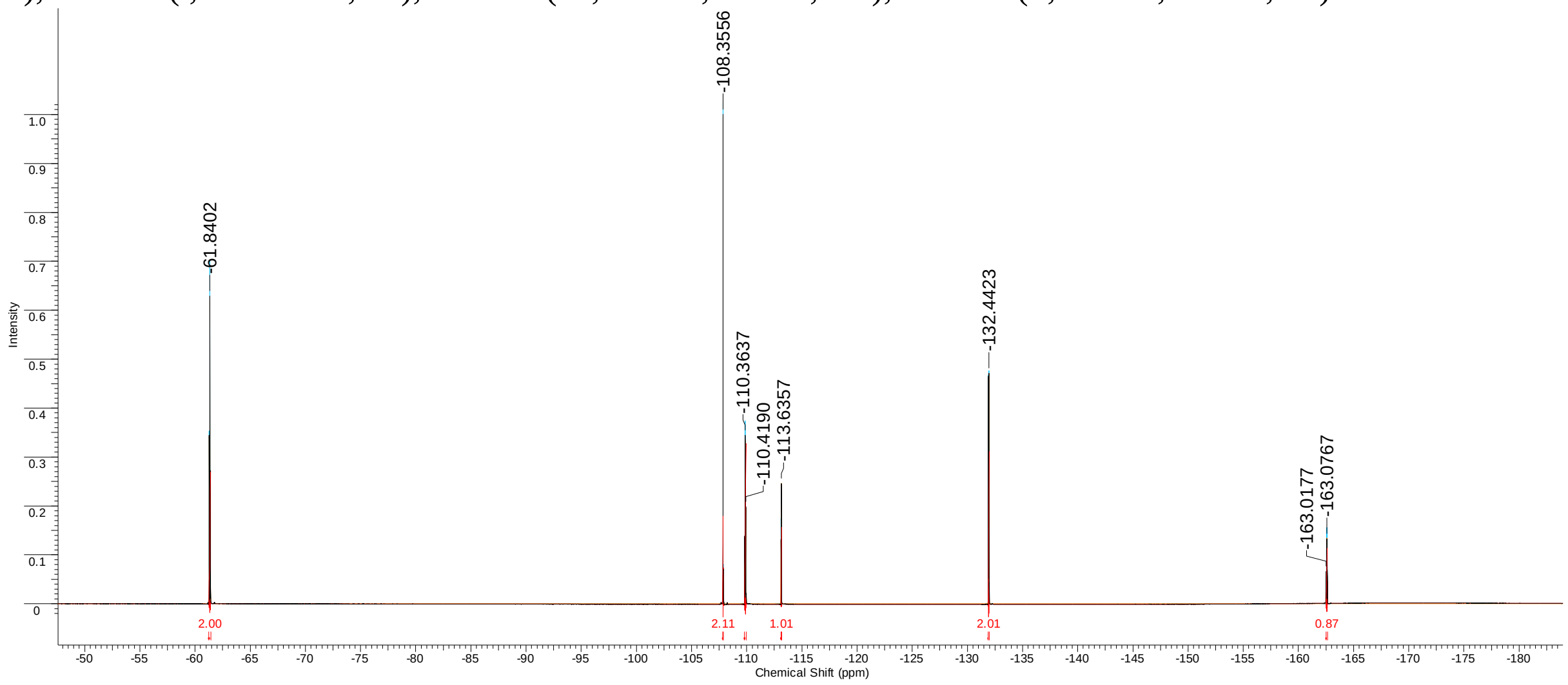

***4'-(difluoro(3,4,5-trifluorophenoxy)methyl)-2,3',5'-trifluoro-[1,1'-biphenyl]-4-yl 2,6-difluoro-4-(5-ethyl-1,3-dioxan-2-yl)benzoate (MUT_JK102)***

Yield 2.0g (68.5%)

Purity 99.4% (99.4/0.2 trans/cis) (HPLC-MS)

MS(EI) m/z: 673 (M+); 655; 587; 569; 527; 457; 441; 419; 384; 272; 255; 244; 169; 141

$^{1}$H NMR (500 MHz, $CDCl_3$) δ: 7.52 (t, *J*=8.55 Hz, 1 H); 7.23 (m, 6 H); 7.02 (dd, *J*=7.63, 5.80 Hz, 2 H); 5.42 (s, 1 H); 4.30 (dd, *J*=11.75, 4.73 Hz, 2 H); 3.57 (t, *J*=11.44 Hz, 2 H); 2.07 (m, 1 H); 1.20 (m, 2 H); 0.97 (t, *J*=7.63 Hz, 3 H)

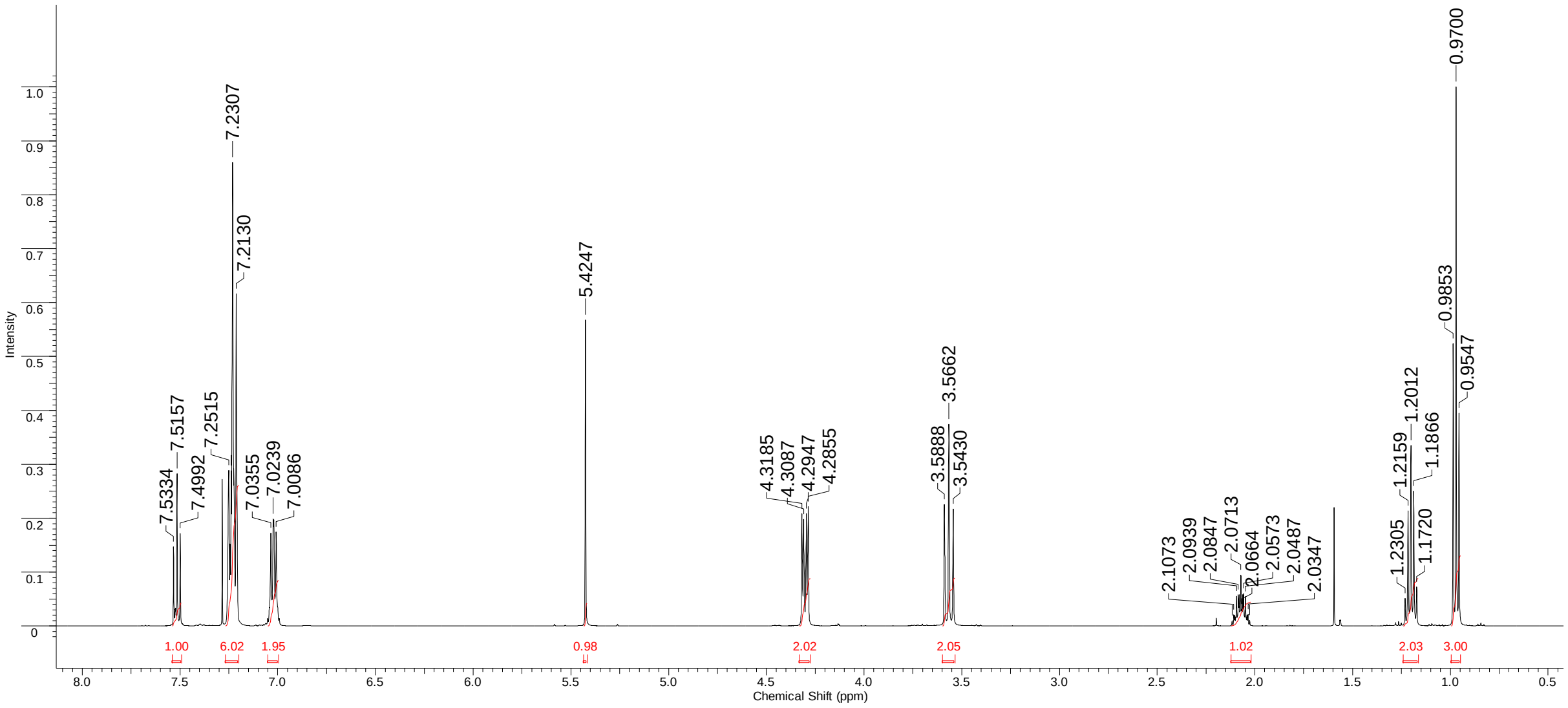

$^{13}C$ NMR (125 MHz, $CDCl_3$) δ: 161.99 (d, *J*=5.45 Hz); 161.00 (d, *J*=8.17 Hz); 160.53; 159.93 (d, *J*=5.45 Hz); 159.20; 158.94 (d, *J*=5.45 Hz); 158.52; 152.03 (q, *J*=10.90, 5.45 Hz); 151.66 (d, *J*=10.90 Hz,); 150.04 (q, *J*=10.90, 5.45 Hz); 145.76 (t, *J*=9.99 Hz); 144.63 (m); 140.69 (t, *J*=10.90 Hz); 138.50 (dt, *J*=250.69, 14.99 Hz); 130.63 (d, *J*=3.63 Hz); 123.64 (d, *J*=12.72 Hz); 120.15 (t, J=266.13 Hz); 118.36 (d, *J*=3.63 Hz); 113.15 (dd, *J*=27.25, 3.63 Hz); 110.95; 110.74; 110.37 (m); 109.24 (m); 107.50 (m); 98.81 (t, J=1.82 Hz); 72.42; 35.74; 21.14; 10.90

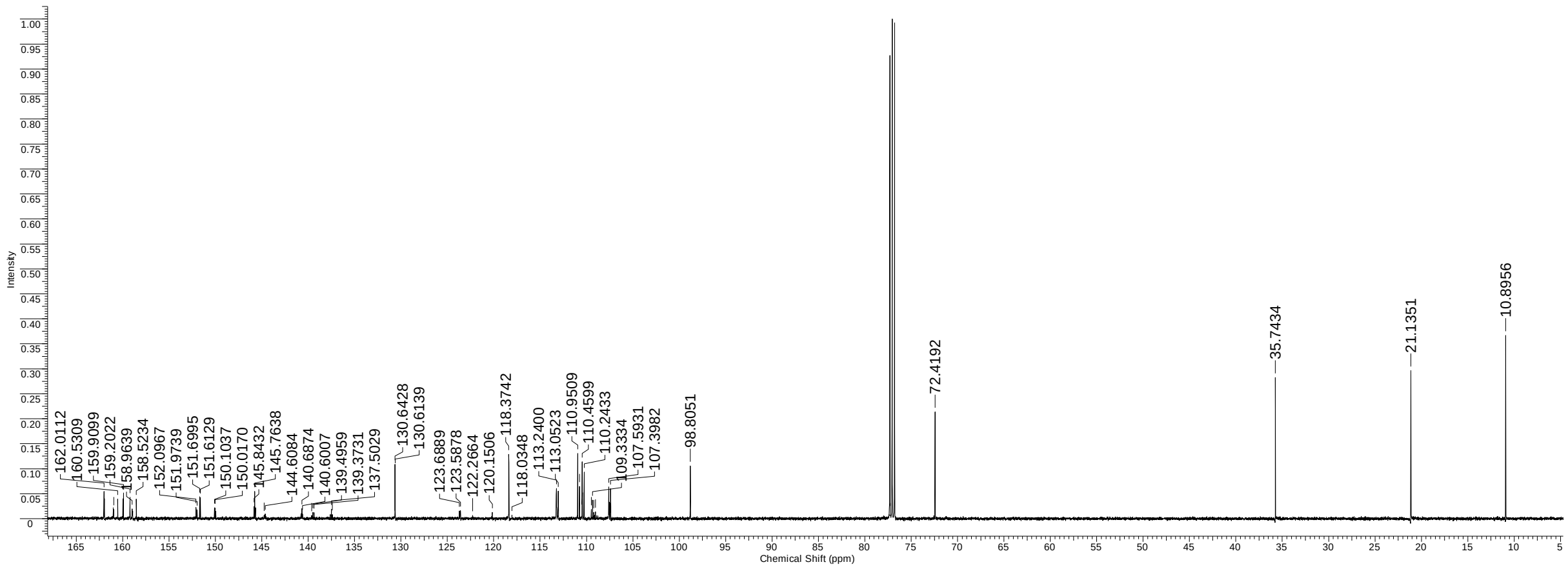

$^{19}F$ NMR (471 MHz, $CDCl_3$) δ: -62.51 (t, J=26.0 Hz, 2 F); -109.05 (d, *J*=10.4 Hz, 2 F); -111.04 (td, J=26.0, 10.4 Hz, 2 F); -114.30 (t, J=10.4 Hz, 1 F); -133.10 (dd, J=21.2; 6.9 Hz, 2 F); -163.75 (tt, J=20.8, 5.2 Hz, 1 F)

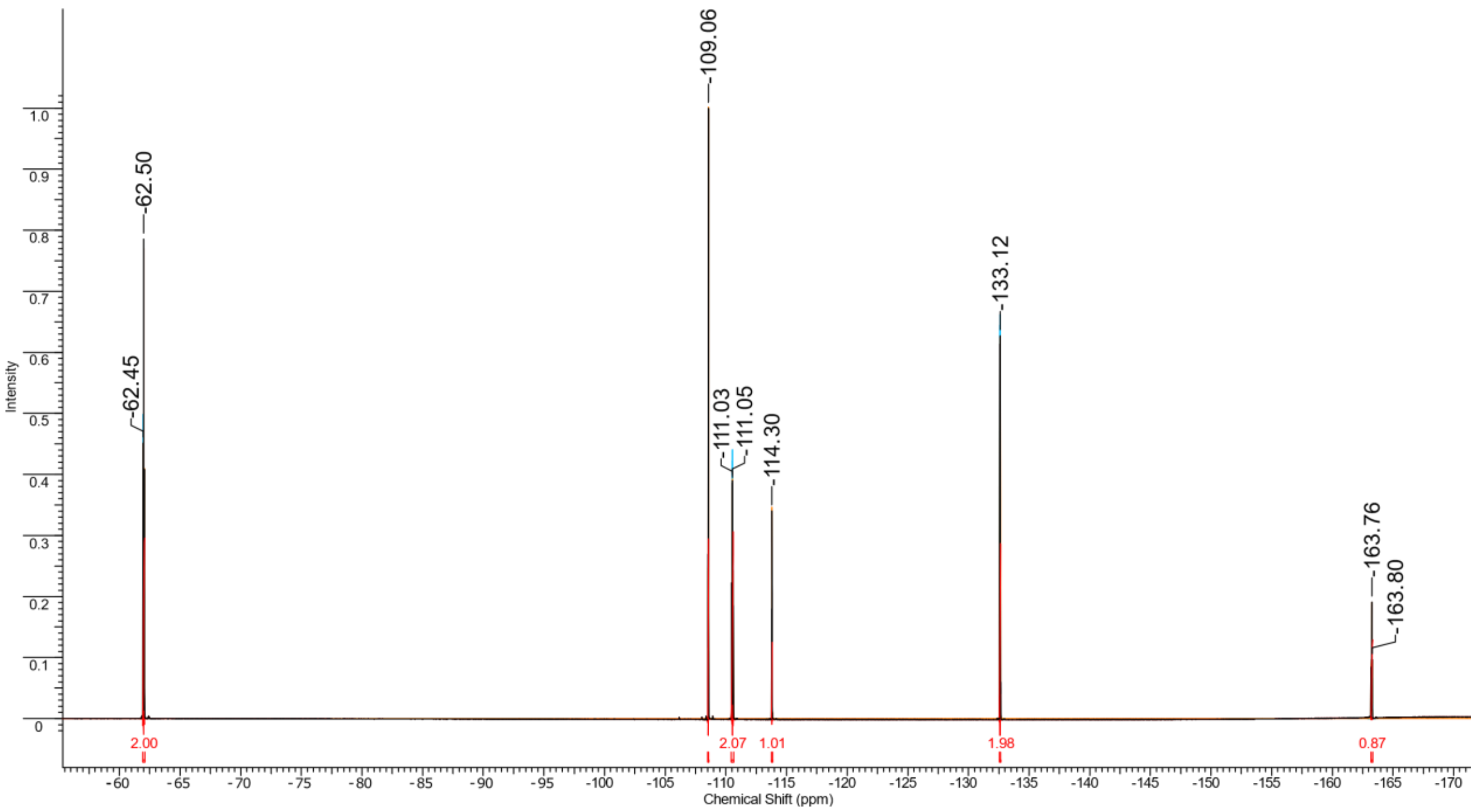

***4'-(difluoro(3,4,5-trifluorophenoxy)methyl)-2,3',5'-trifluoro-[1,1'-biphenyl]-4-yl 2,6-difluoro-4-(5-butyl-1,3-dioxan-2-yl)benzoate (MUT_JK104)***

Yield 1.7g (50.0%)

Purity 99,9% (99.9/0.1 trans/cis) (HPLC-MS)

MS(EI) m/z: 702 (M+); 684; 607; 587; 569; 555; 545; 457; 441; 419; 384; 283; 272; 244; 169; 141

$^1$H NMR (500 MHz, $CDCl_3$) δ: 7.52 (t, *J*=8.70 Hz, 1 H); 7.23 (m, 6 H); 7.02 (dd, *J*=7.48, 5.95 Hz, 2 H); 5.43 (s, 1 H); 4.28 (dd, *J*=11.90, 4.58 Hz, 2 H); 3.56 (t, *J*=11.44 Hz, 2 H); 2.14 (m, 1 H); 1.33 (m, 4 H); 1.14 (m, 2 H); 0.94 (t, *J*=7.02 Hz, 3 H)

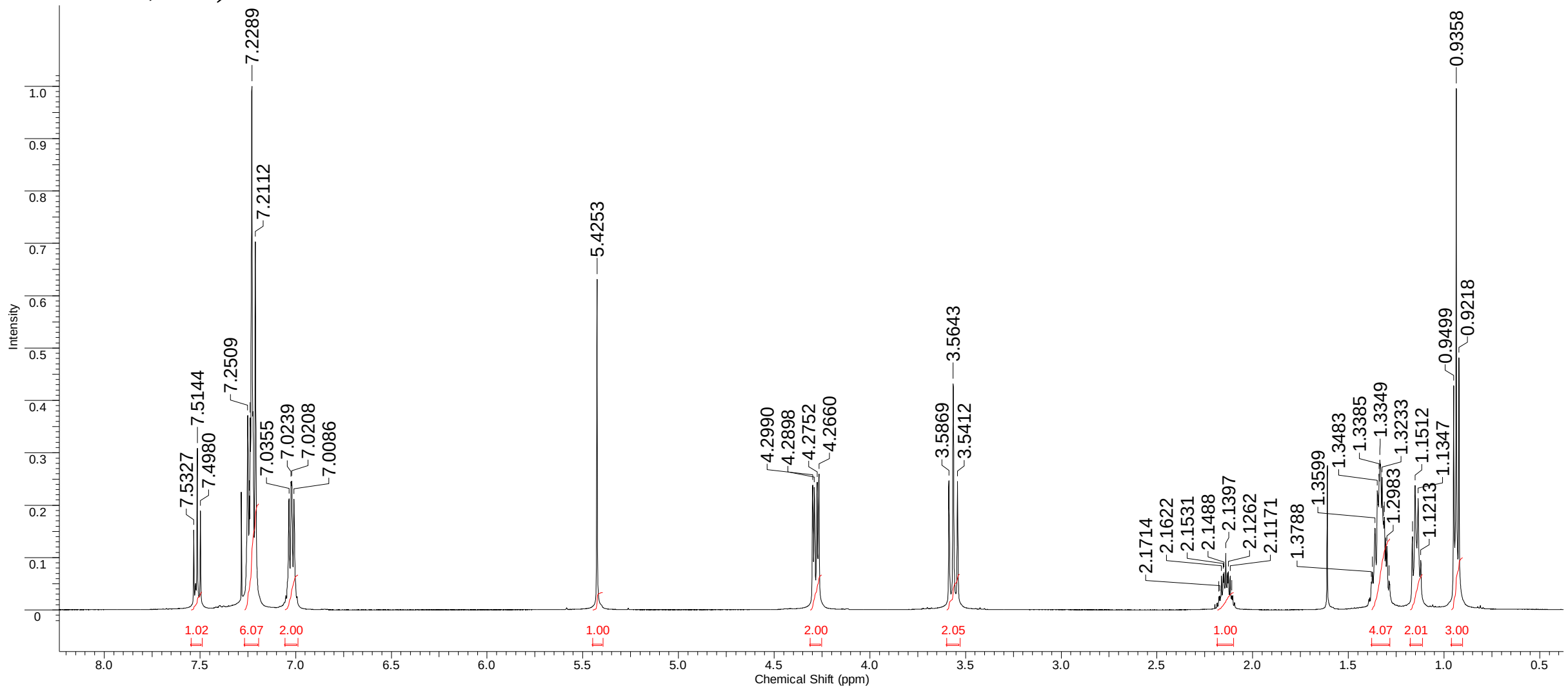

$^{13}$C NMR (125 MHz, $CDCl_3$) δ: 161.99 (d, *J*=5.45 Hz); 160.99 (d, *J*=5.45 Hz); 160.53; 159.93 (d, *J*=5.45 Hz); 159.20; 158.94 (d, *J*=6.36 Hz); 158.52; 152.03 (q, *J*=10.90, 5.45 Hz); 151.66 (d, *J*=10.90 Hz); 150.04 (q, *J*=10.45, 5.00 Hz); 145.77 (t, *J*=9.99 Hz); 144.63 (m); 140.69 (t, *J*=10.90 Hz); 138.49 (dt, J=249.77, 15.45 Hz); 130.63 (d, *J*=3.63 Hz); 123.63 (d, *J*=12.72 Hz); 120.15 (t, J=267.03 Hz); 118.36 (d, *J*=3.63 Hz); 113.14 (dd, *J*=27.70, 3.18 Hz); 110.95; 110.74; 110.35 (m); 109.26 (m); 107.49 (m); 98.80 (t, *J*=1.82 Hz); 72.64; 34.15; 28.49; 27.80; 22.83; 13.89

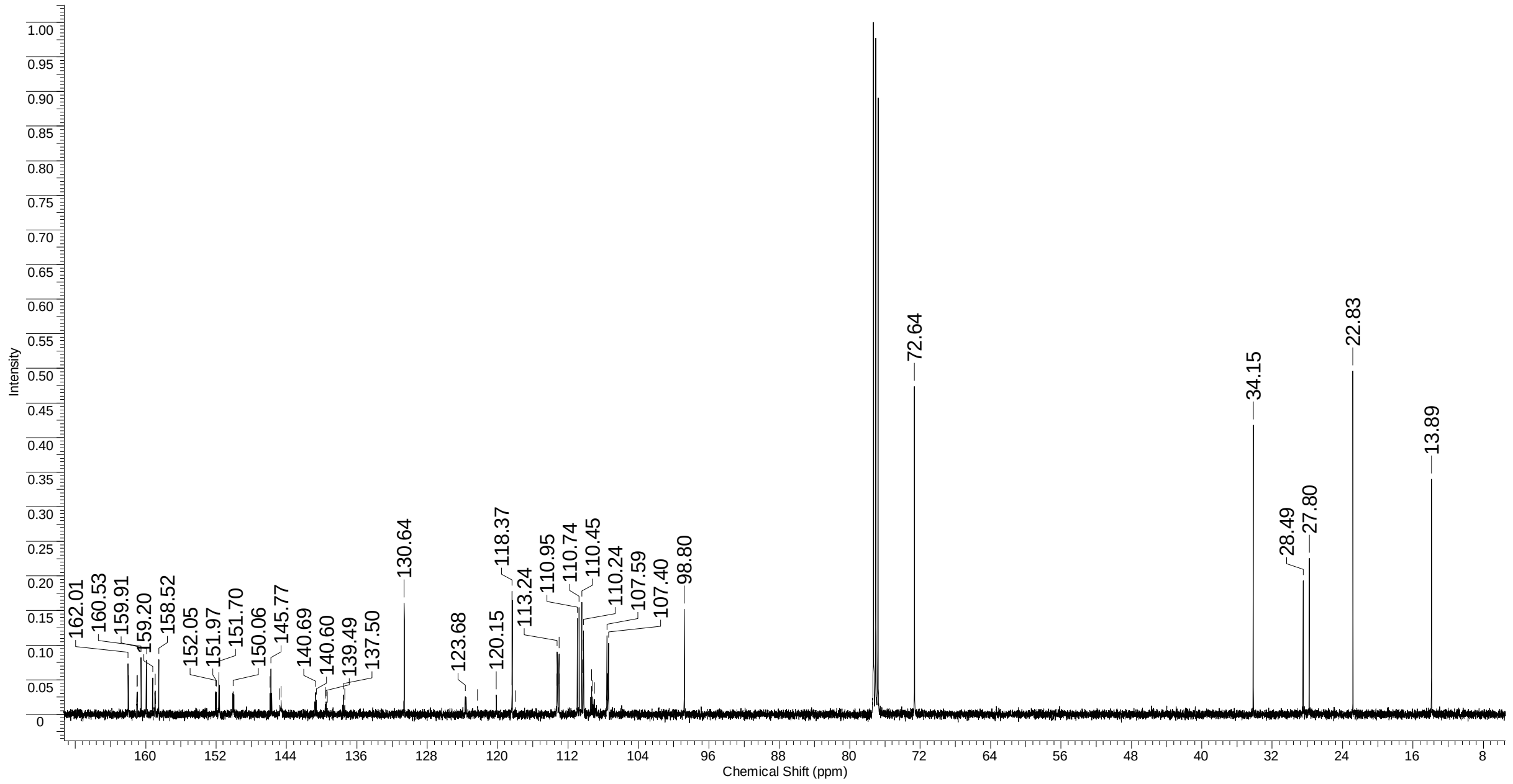

$^{19}$F NMR (471 MHz, $CDCl_3$) δ: -62.51 (t, *J*=26.01 Hz, 2 F); -109.04 (d, *J*=8.67 Hz, 2 F); -111.04 (td, *J*=26.01, 10.41 Hz, 2 F); -114.31 (t, J=10.4 Hz, 1 F); -133.10 (dd, *J*=20.80, 8.67 Hz, 2 F); -163.76 (tt, *J*=20.81, 5.20 Hz, 1 F)

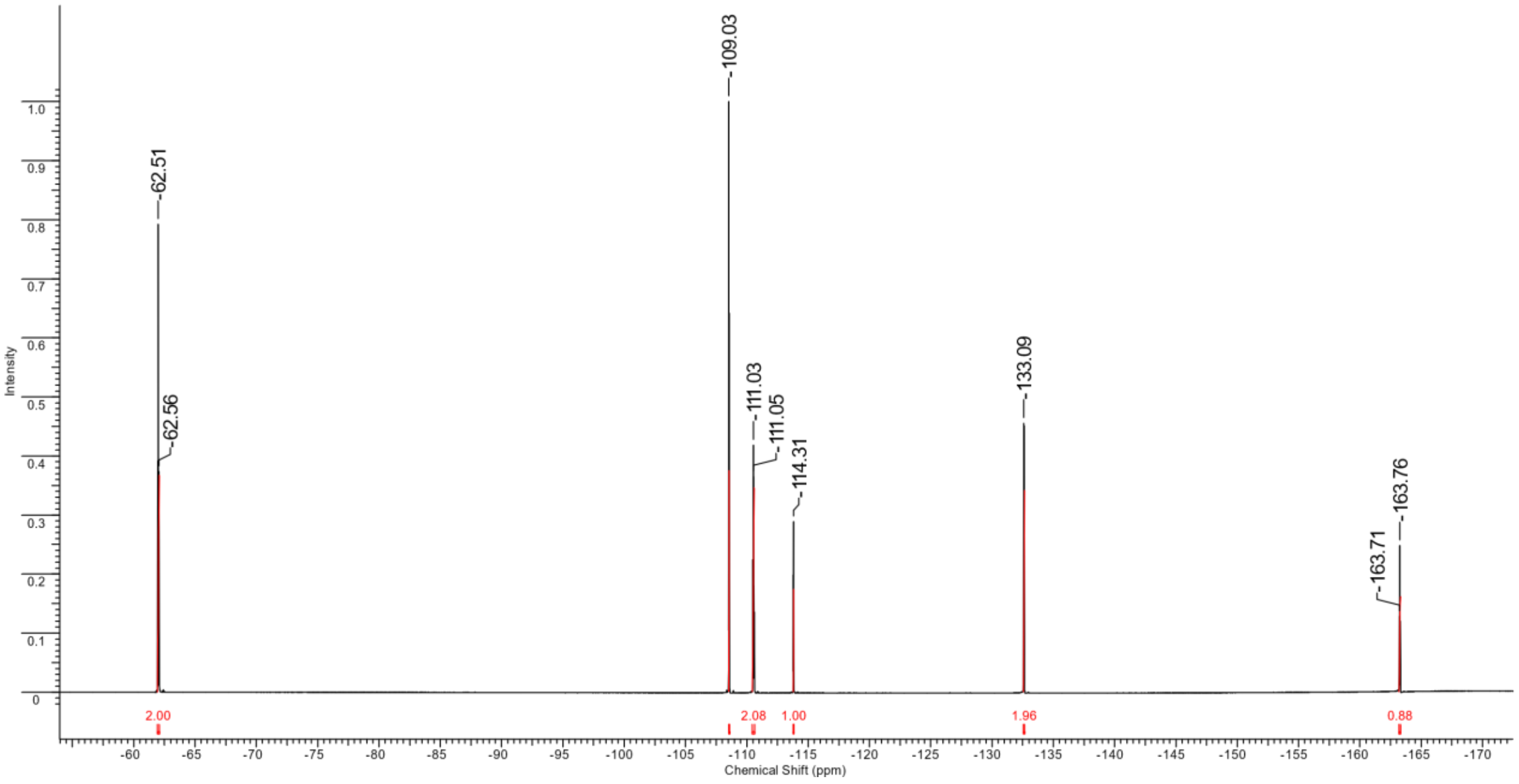

***4'-(difluoro(3,4,5-trifluorophenoxy)methyl)-2,3',5'-trifluoro-[1,1'-biphenyl]-4-yl 2,6-difluoro-4-(5-pentyl-1,3-dioxan-2-yl)benzoate (MUT_JK105)***

Yield 1.9g (54.6%)

Purity 99.2% (99.2/0.3 trans/cis) (HPLC-MS)

MS(EI) m/z; 715 (M+); 697; 587; 569; 541; 441; 413; 384; 297;272; 244; 169; 141

$^{1}$H NMR (500 MHz, $CDCl_3$) δ: 7.51 (t, *J*=8.70 Hz, 1 H); 7.23 (m, 6 H); 7.02 (dd, *J*=7.63, 6.10 Hz, 2 H); 5.42 (s, 1 H); 4.28 (dd, *J*=11.75, 4.73 Hz, 2 H); 3.56 (t, *J*=11.44 Hz, 2 H); 2.14 (m, 1 H); 1.32 (m, 6 H); 1.13 (m, 2 H); 0.92 (m, 3 H)

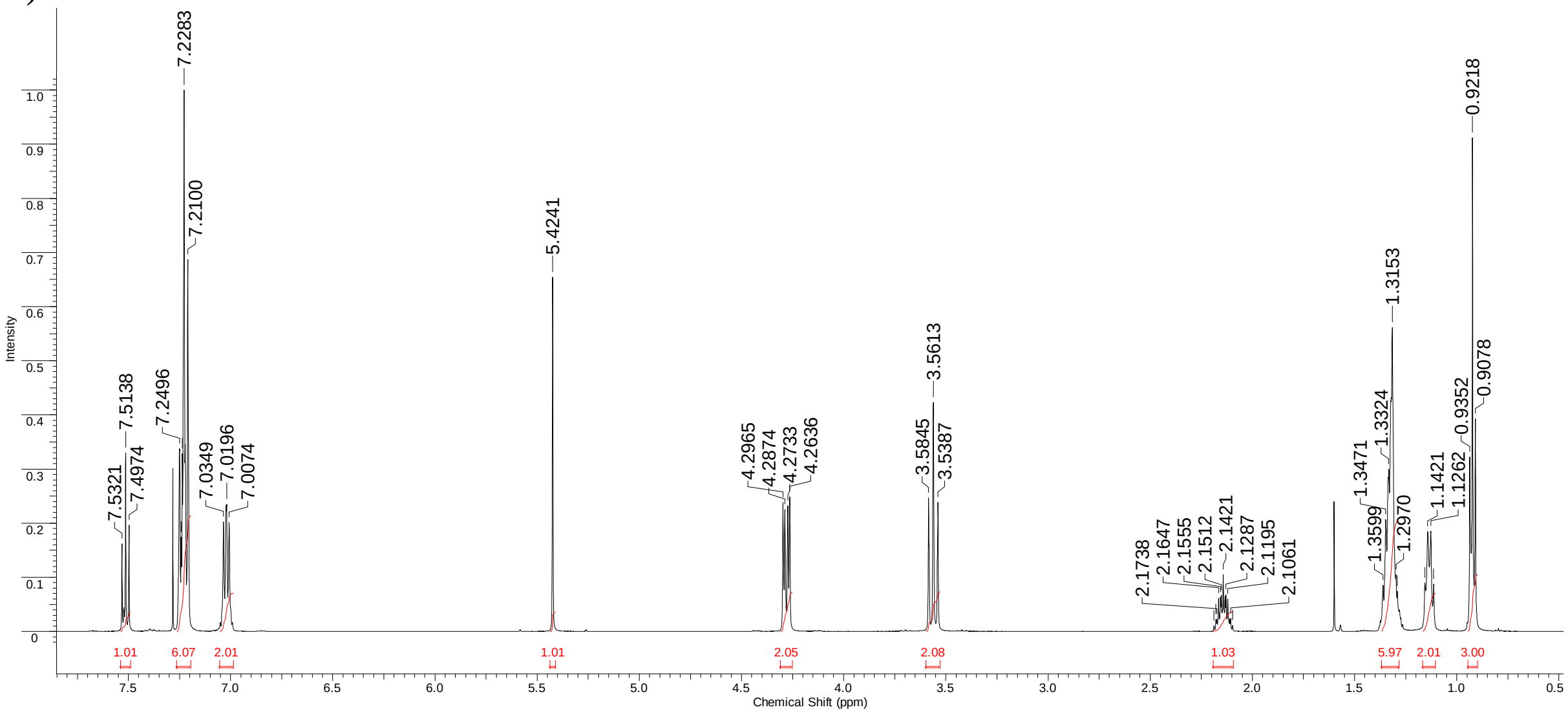

$^{13}$C NMR (125 MHz, $CDCl_3$) δ: 161.99 (d, *J*=5.45 Hz); 160.99 (d, *J*=5.45 Hz); 160.53; 159.93 (d, *J*=5.45 Hz); 159.20; 158.94 (d, *J*=6.36 Hz); 158.52; 152.03 (q, *J*=10.90, 4.54 Hz); 151.65 (d, *J*=10.90 Hz); 150.04 (q, *J*=10.90, 4.54 Hz); 145.77 (t, *J*=9.99 Hz); 144.61 (m); 140.68 (t, *J*=10.90 Hz, 1 C); 138.49 (dt, *J*=250.46, 14.99 Hz); 130.63 (d, *J*=3.63 Hz); 123.64 (d, *J*=12.72 Hz); 120.15 (t, J=266.13 Hz); 118.36 (d, *J*=3.63 Hz); 113.14 (dd, J=27.5, 3.64 Hz); 110.95; 110.75; 110.35 (m); 109.26 (m); 98.80 (t, J=1.81 Hz); 72.64; 34.15; 31.93; 28.05; 25.99; 22.49; 14.02

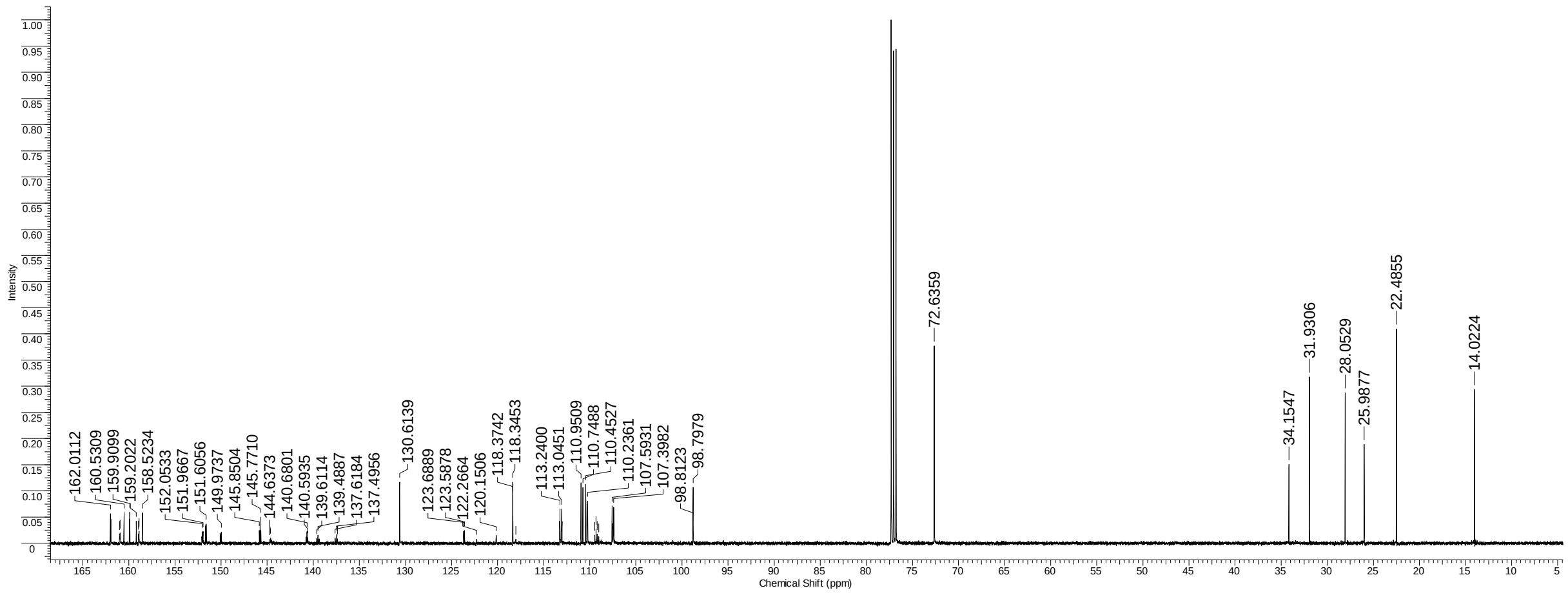

$^{19}$F NMR (471 MHz, $CDCl_3$) δ: -62.51 (t, *J*=26.01 Hz, 2 F); -109.04 (d, *J*=8.67 Hz, 2 F); -111.04 (td, *J*=26.01, 10.41 Hz, 2 F); -114.30 (t, J=10.4 Hz, 1 F); -133.10 (dd, J=20.80, 8.67 Hz, 2 F); -163.75 (tt, J=20.81, 5.20 Hz, 1 F)

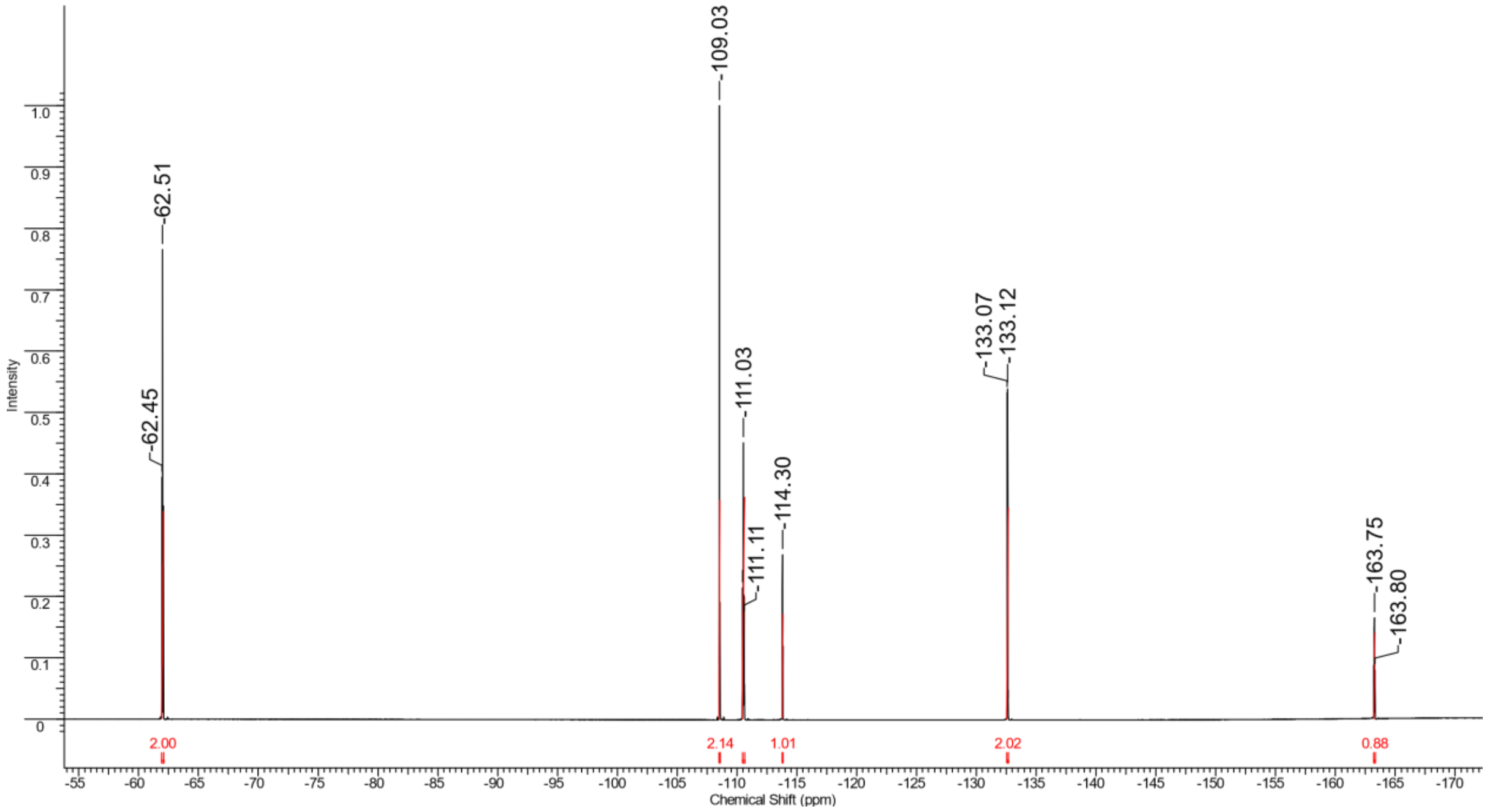

## Supplementary Text

**Table S1**. Phase sequence, phase transition temperatures (in °C) and associated thermal effects (in parentheses, kJ mol$^{-1}$) for compounds **MUT_JK10n**, obtained from DSC studies. Values were extracted from the second heating scan, except for $N_{TBF}$ - $SmC_F$ phase transition in homolog **MUT_JK102** - due to the monotropic nature of the phases a cooling run was analyzed.

| *Compound* | **n** | *Phase sequence* |
|---|---|---|
| **MUT_JK101** | 1 | Cr 114.2 (31.30) [$N_{TBF}$ 78.4[a]] $N_F$ 190.2 (0.99) N 217.7 (1.32) Iso |
| **MUT_JK102** | 2 | Cr 98.0 (28.87) [$SmC_F$ 82.1 (0.23) $N_{TBF}$ 97.3[a]] $N_F$ 161.5 (0.48) N 223.9 (1.23) Iso |
| **MUT_JK103** | 3 | Cr 67.4 (20.24) $SmC_F$ 90.9 (0.06) $N_{TBF}$ 103.3 (0.0007) $N_F$ 143.1 (0.28) $N_X$ 150.7 (0.02[b]) N 234.1 (1.10) Iso |
| **MUT_JK104** | 4 | Cr 72.3 (31.01) $SmC_F$ 103.0 (0.17) $SmA_F$ 125.2 (0.40) SmA 143.5 (0.11) N 231.2 (1.26) Iso |
| **MUT_JK105** | 5 | Cr 57.4 (31.70) $SmC_F$ 86.0 (1.30) SmA 163.7 (0.09) N 229.8 (1.39) Iso |

[a] from microscopic observations
[b] heat capacity jump, in kJ mol$^{-1}$ K$^{-1}$

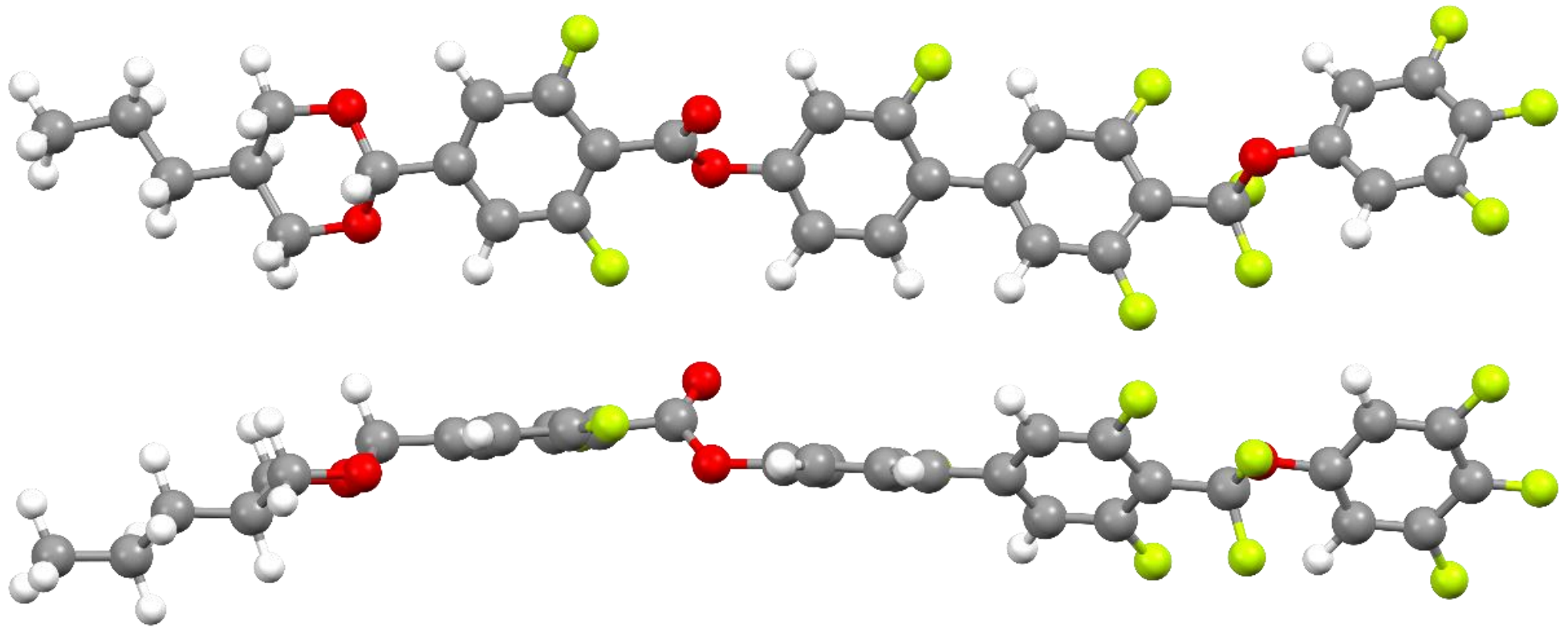

**Figure S2.** Optimized molecular structure of **MUT_JK103**.

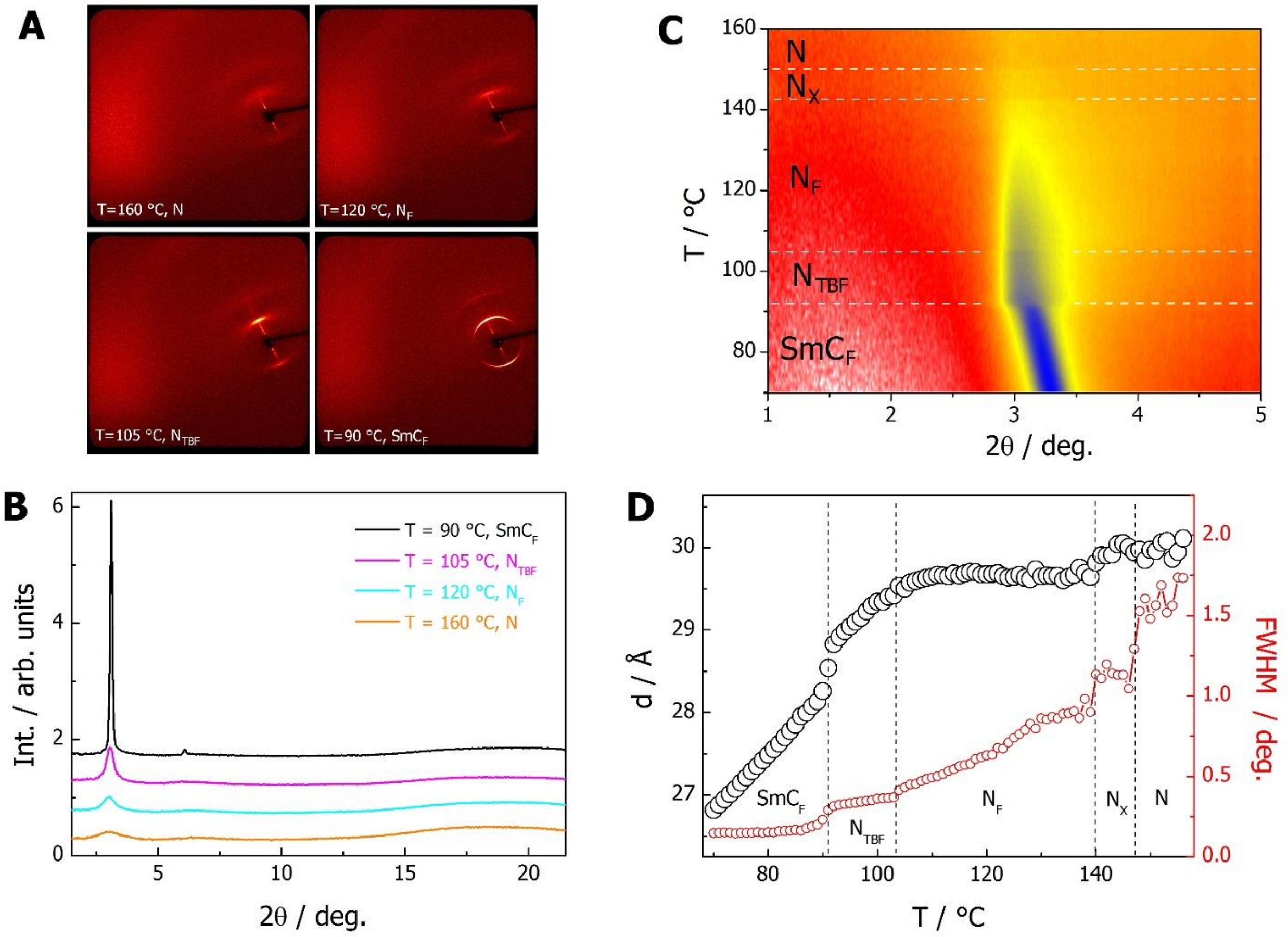

**Figure S3**. **A** Broad-angle 2D XRD patterns taken for a partially aligned sample of **MUT_JK103** in the N, $N_F$, $N_{TBF}$ and $SmC_F$ phases, at the indicated temperatures. **B** X-ray diffractograms obtained by integration of 2D patterns given in **A**, curves were vertically shifted for clarity of presentation. **C** Temperature evolution of the low-angle diffraction pattern, recorded across the N-$N_F$-$N_{TBF}$-$SmC_F$ phase sequence. **D** Layer spacing in the smectic phase and local periodicity of the structure in nematic phases (black circles), and the full width at half maximum (FWHM, red circles) of the related diffraction signal, determined from the data presented in **C**. Only in the lowest temperature LC phase the low-angle diffraction signal becomes of machine-resolution, proving the long-range positional order of molecules in the smectic phase. In all other LC phases the signal is additionally broadened, due to the finite value of the correlation length of positional order, thus the phases are identified as nematics. Gradual narrowing of the signal on cooling, with step-like changes at the phase transitions, evidences an increase of the correlation length. In the lowest temperature nematic phase, $N_{TBF}$, it reaches 100 Å.

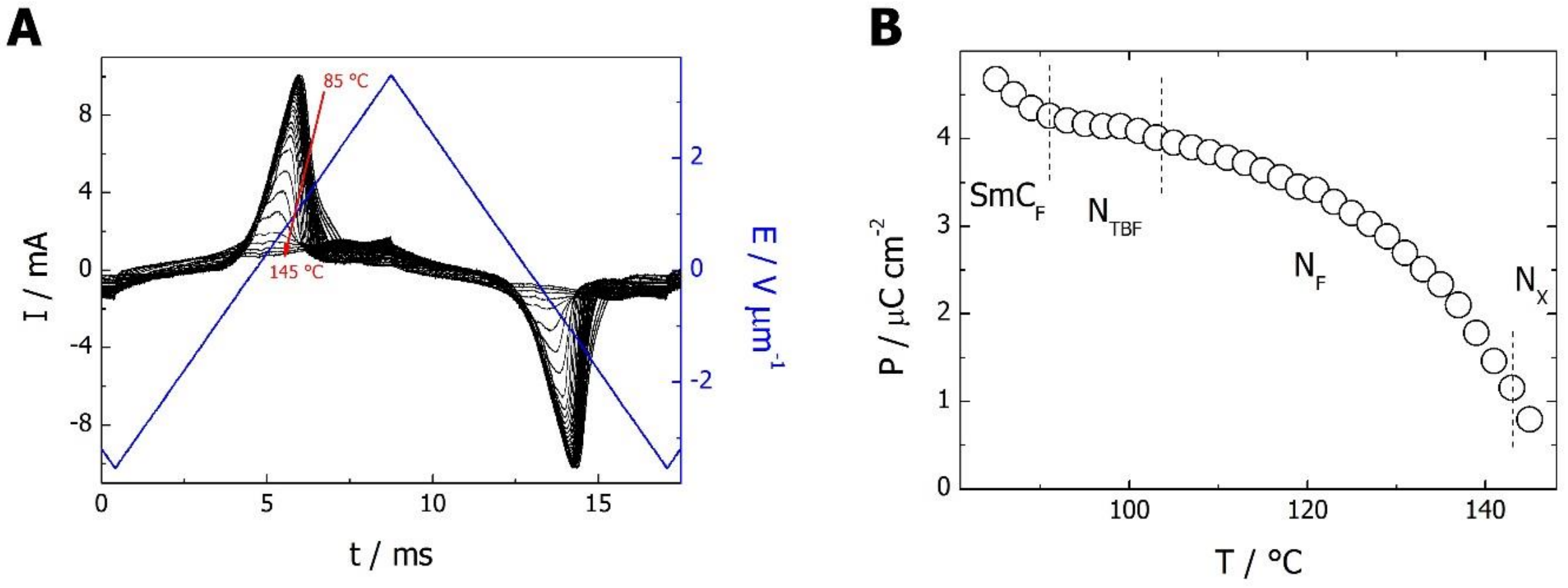

**Figure S4**. **A** Repolarization current recorded under the application of triangular wave voltage at 60 Hz to the 10-μm-thick cell, measurements were performed on heating. **B** Polarization values vs. temperature obtained by integration of the current peak shown in panel **A**. Magnitude of the polarization, ~4 μC $cm^{-2}$ recorded in the $N_{TBF}$ phase is comparable to that of other ferroelectric nematic materials.

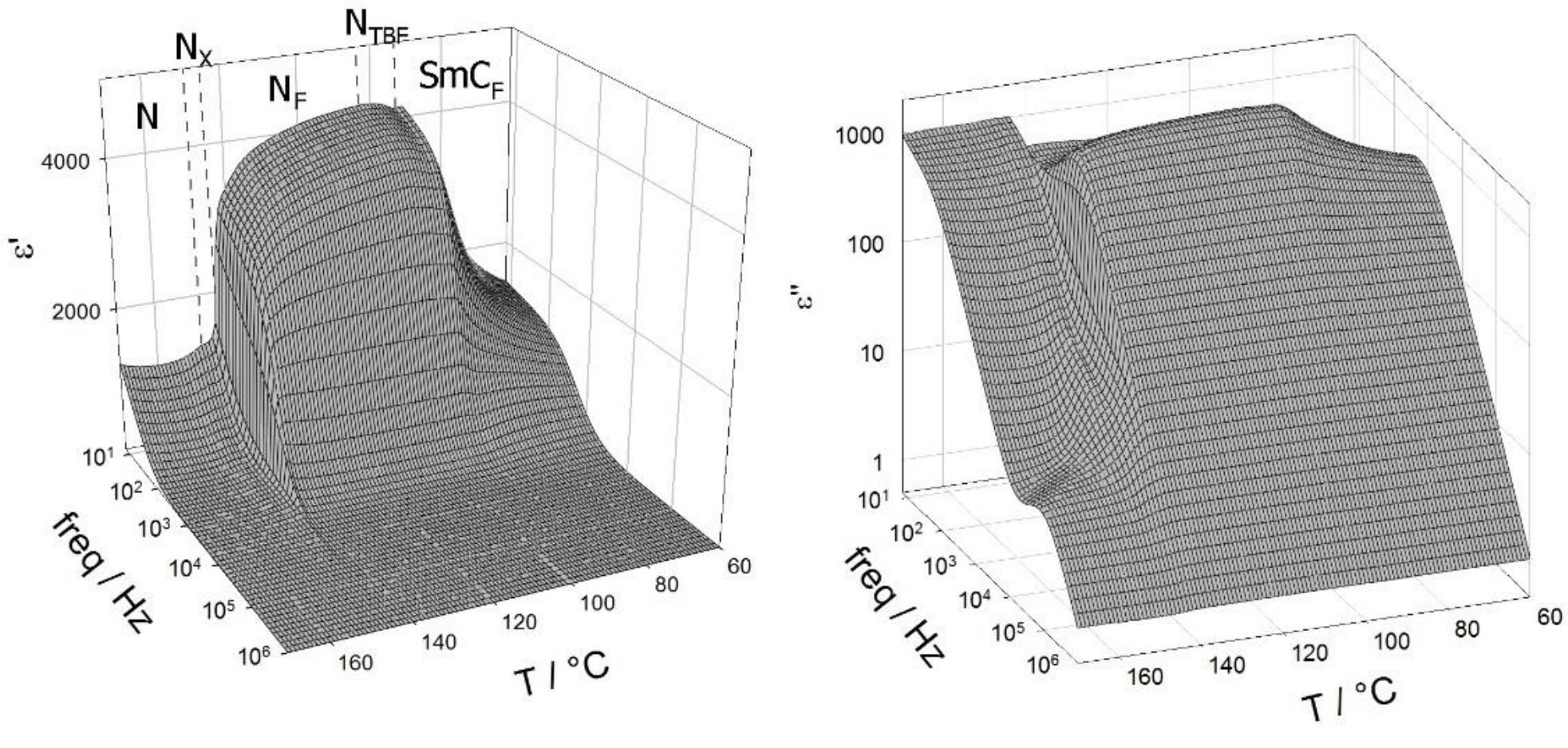

**Figure S5**. Real (ε') and imaginary (ε'') parts of permittivity recorded for compound **MUT_JK103** in 3-μm-thick cell with bare ITO electrodes (without aligning polymer layers). Only small changes are visible at $N_F$-$N_{TBF}$ phase transitions, apparently both phases are ferroelectric, with very similar polar properties. In $SmC_F$ phase the dielectric permittivity decreases gradually.

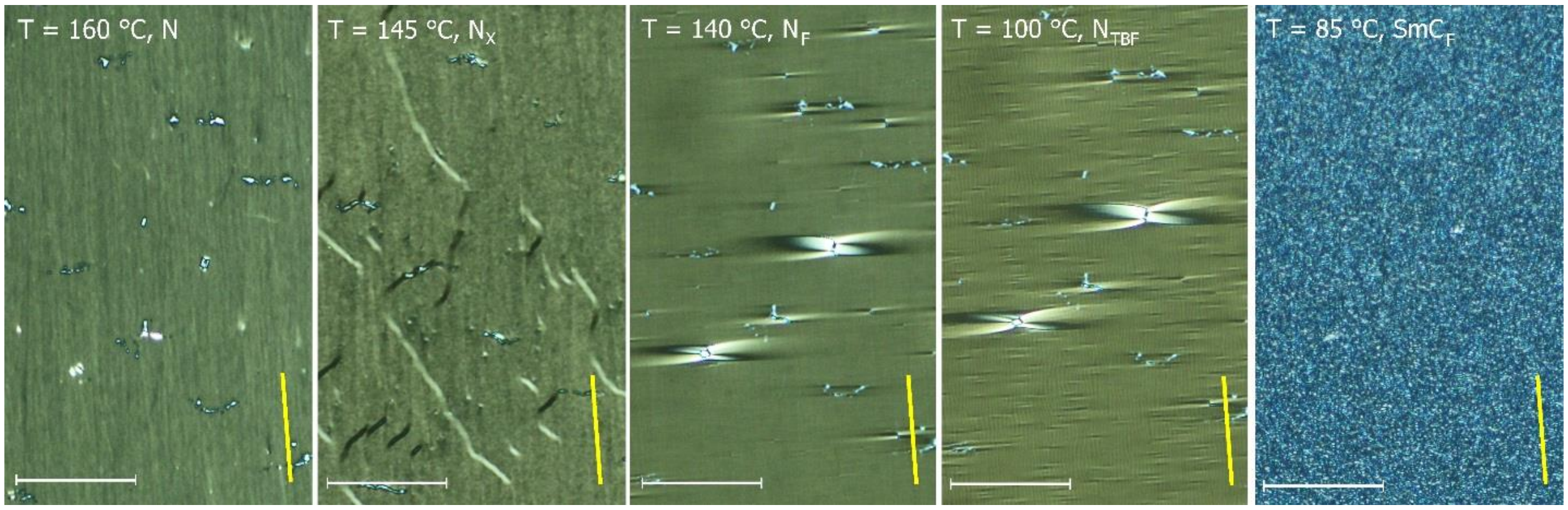

**Figure S6**. Optical textures observed in four nematic phases (N, $N_x$ , $N_F$ and $N_{TBF}$) and smectic $C_F$ phase of compound **MUT_JK103** at indicated temperatures. The sample was prepared in 1.5-μm-thick cell with planar anchoring and parallel rubbing on both surfaces. Images were recorded between crossed polarizers (at horizontal and vertical directions), the rubbing direction (yellow line) was slightly inclined from the vertical direction and scale bars correspond to 100 μm. Chevron defects visible in the second panel allowed for the identification of $N_x$ phase.

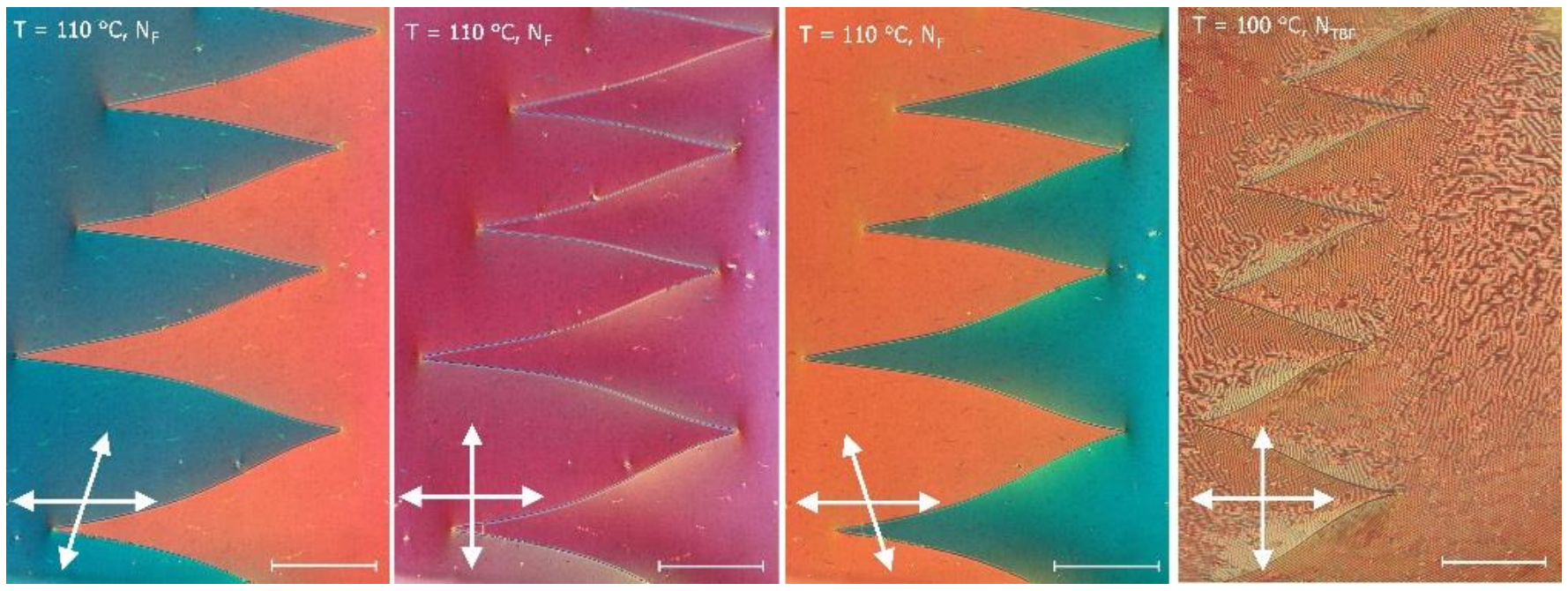

**Figure S7**. Optical textures observed in the $N_F$ and $N_{TBF}$ phases of compound **MUT_JK103**. The sample was prepared in 5-μm-thick cell with planar anchoring and antiparallel rubbing on both surfaces. Images were recorded either between crossed polarizers or after slight de-crossing of polarizers in opposite directions (polarizers orientation marked by arrows). Visible optical activity is due to twisted (TN) states induced by the antiparallel orientation of molecules at both surfaces. Scale bars correspond to 200 μm.

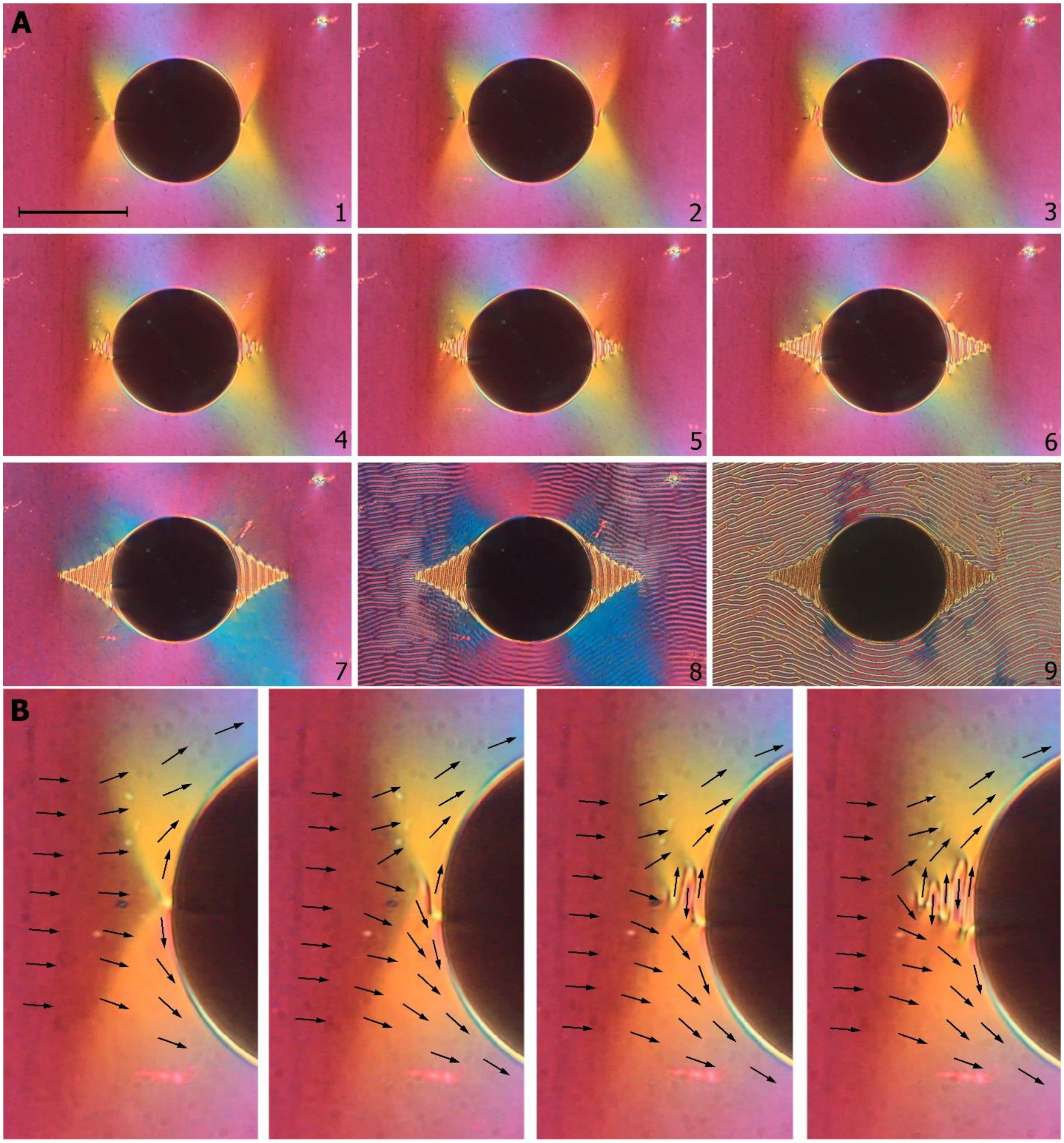

**Figure S8**. **A** Evolution of the optical texture at the $N_F$-$N_{TBF}$ phase transition: in $N_F$ phase the air bubble causes bending of director thus electric polarization field (*41*), formation of the stripe pattern in $N_{TBF}$ starts from half point defects anchored at the air bubble. **B** Enlarged part of images given in **A**, with indicated possible arrangement of polarization vectors in the sample – defect line grows in such a way that finally a regular array of stripes is formed, in which polarization vectors have opposite direction.

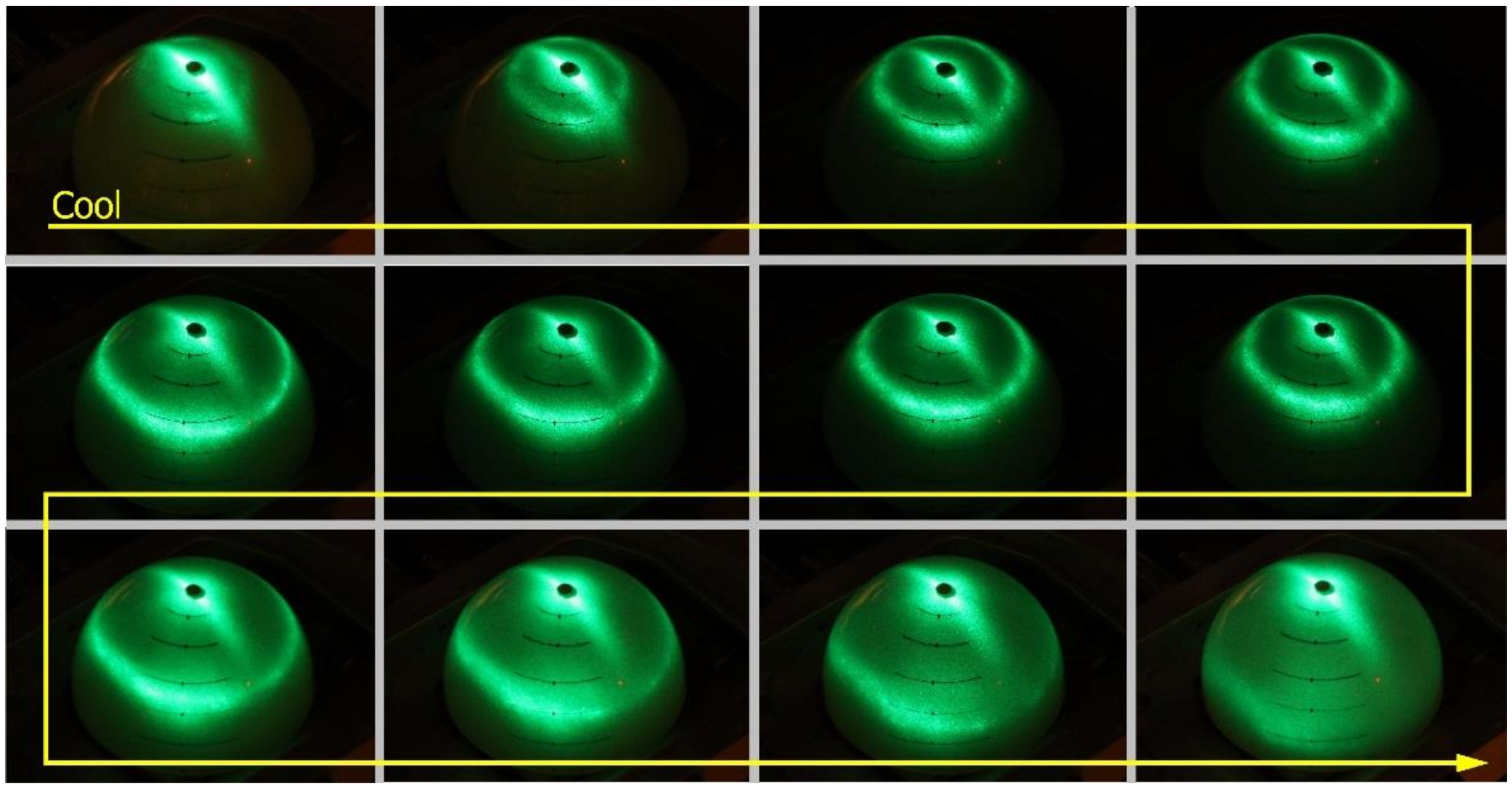

**Figure S9.** Series of images taken during light diffraction experiment performed on the powder-like sample of **MUT_JK103** (one-surface-free droplet of material on a heated surface, illuminated from the bottom with a green laser, diffraction pattern observed on a spherical screen) during cooling through the upper temperature range (5 K) of $N_{TBF}$ phase. Strong changes in diffraction signal position evidence rapid shortening of the pitch of the heliconical structure with decreasing temperature.

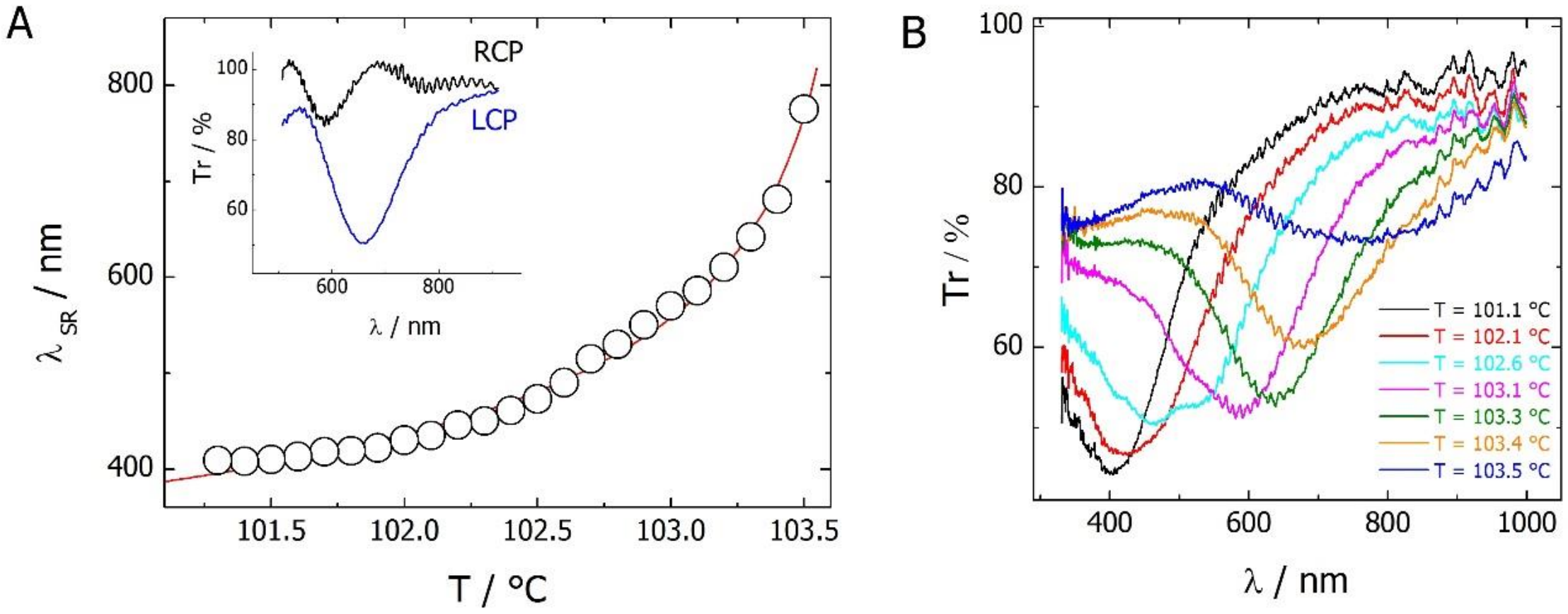

**Figure S10. A** Selective reflection wavelength ($\lambda_{SR}$) measured vs. temperature by scanning the transmission of light through a 10-μm-thick cell with no aligning layers, the selective reflection is blue shifted compared to the helical pitch length determined in light diffraction experiments – apparently the helices are inclined to the light propagation direction. Selective reflection was found to be sensitive to the handedness of circular polarization of light, when measured in micron-sized areas the transmitted light intensity shows a clear minimum only for one of the circular polarizations (data in inset, recorded at T=103.3 °C), depending on the place in the sample. **B** Transmission spectra recorded at indicated temperatures, showing changes in the position and shape of the minimum related to selective reflection of light due to helical structure.

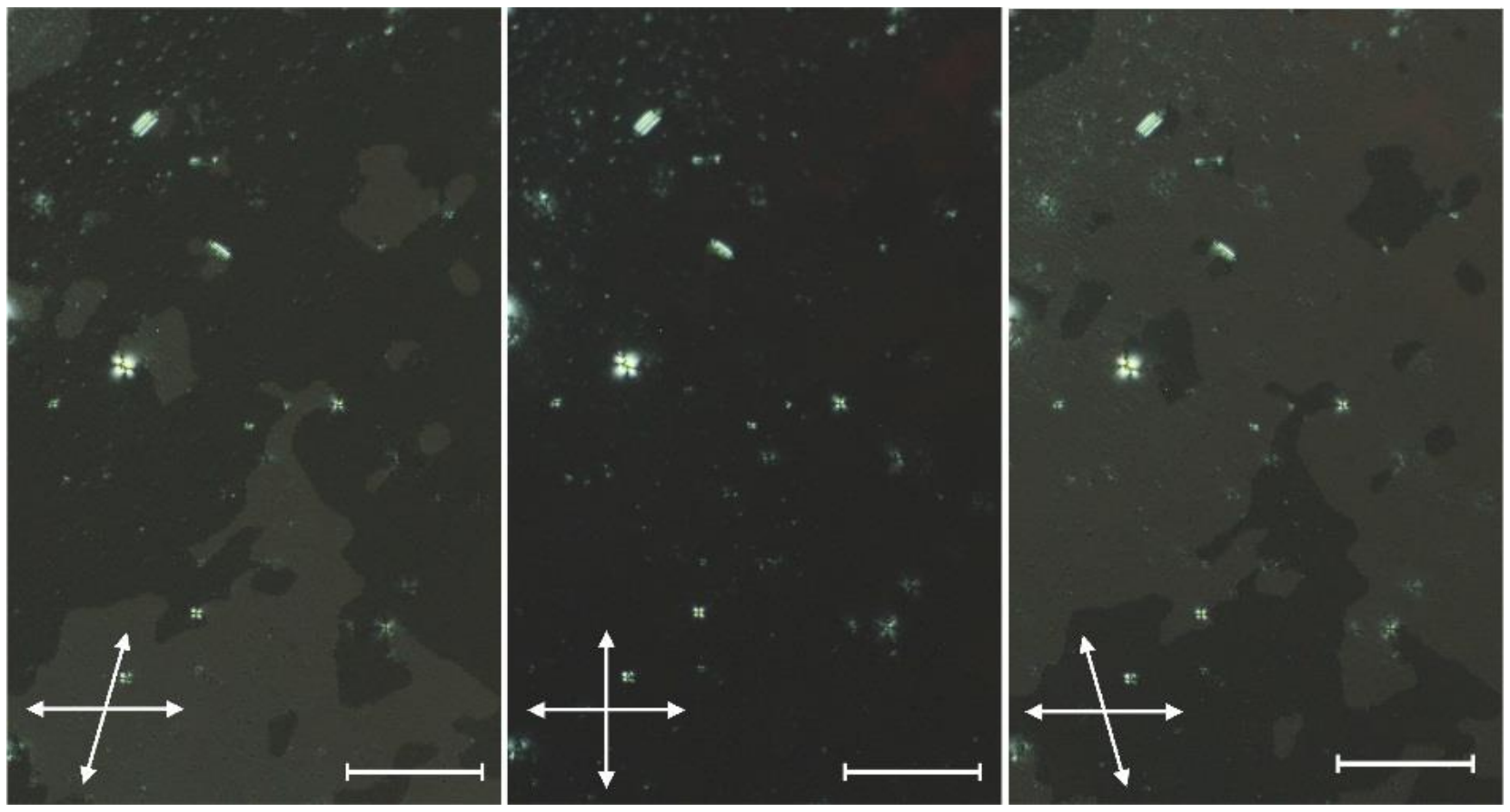

**Figure S11**. Non-birefringent texture of $N_{TBF}$ phase (T = 97 °C) with optically active domains (clearly visible by un-crossing of polarizers, arrows), recorded in a 5-μm-thick cell with planar anchoring under application of weak dc electric field, 0.25 V $\mu m^{-1}$. The electric field causes realignment of the sample - the helix becomes oriented along the light propagation direction. The domains of opposite optical activity correspond to areas in which the helix has opposite handedness. Scale bars correspond to 100 μm.

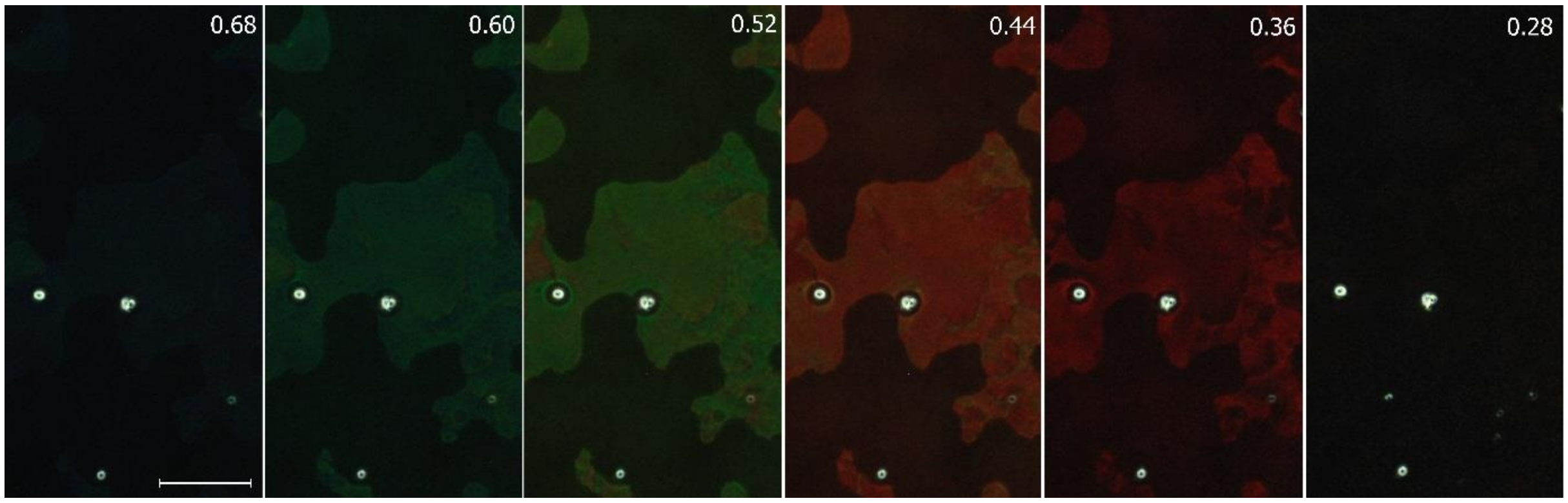

**Figure S12**. Microscopic images presenting selective reflection of light in the $N_{TBF}$ phase (T = 97 °C), they were taken for a 5-μm-thick cell with no alignment layers under application of dc electric field (indicated in upper right corners, in V $\mu m^{-1}$). Two types of domains are visible, showing opposite signs of optical activity due to opposite handedness of the helix, one type of the domains is brought into light extinction condition by slightly de-crossing polarizers. The wavelength of reflection band strongly changes through the whole visible range with applied electric field – apparently the helical pitch gets shorter with increasing of applied voltage. Scale bar corresponds to 100 μm.

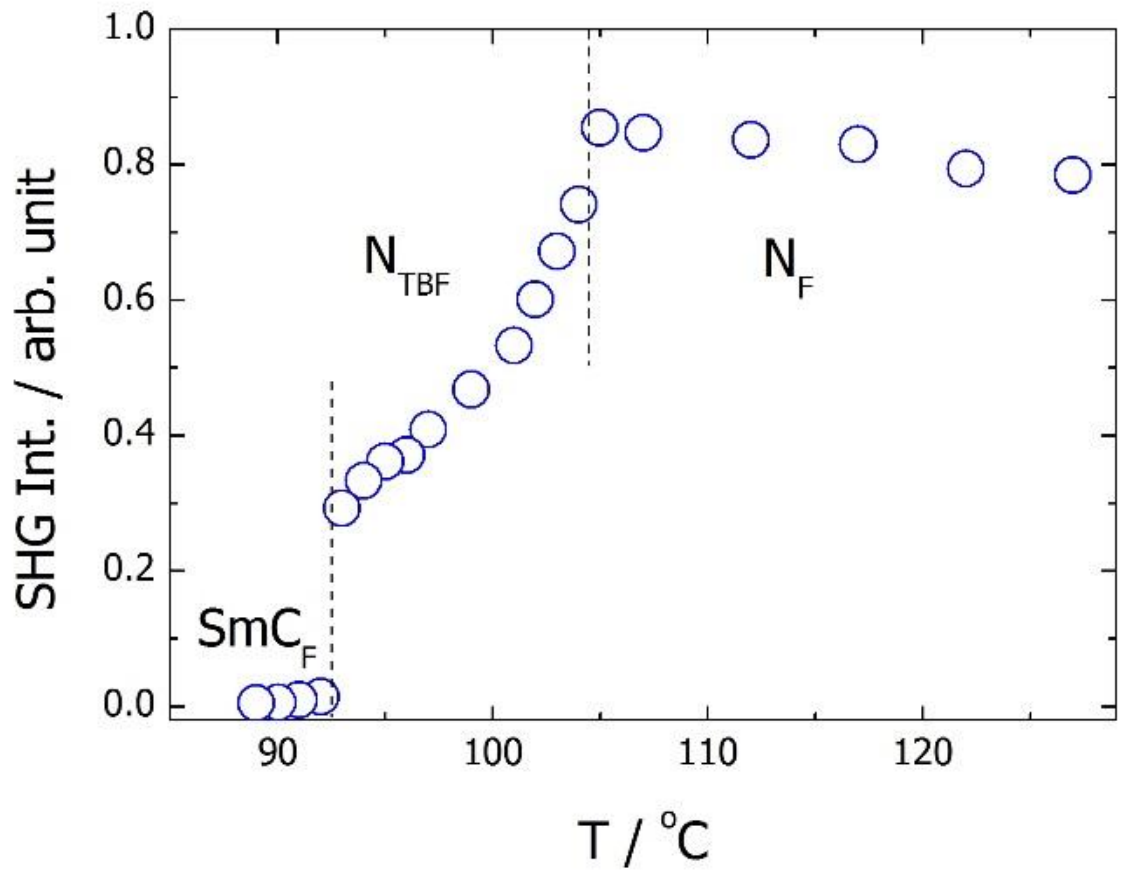

**Figure S13**. Intensity of SHG signal (in arbitrary units) measured vs. temperature for compound **MUT_JK103**. The decrease of SHG in the $N_{TBF}$ phase relative to the $N_F$ phase is related to the partial compensation of spontaneous electric polarization due to the formation of heliconical structure. In the $SmC_F$ phase the SHG signal is much weaker, although not zero. The phase is obviously SHG active, however due to the very grainy texture formed in this phase there is strong light scattering which drastically reduces the measured SHG signal.